\documentclass[nojss]{jss}
\usepackage{soul} 

\usepackage{subcaption}

\usepackage{amsmath}
\usepackage{enumitem}

\usepackage{orcidlink,thumbpdf,lmodern}

\usepackage{framed}

\author{Gian Pietro Bellocca~\orcidlink{0009-0006-3036-5416}\\Universidad Carlos III de Madrid
   \And Ignacio Garrón~\orcidlink{0000-0002-8232-0029}\\Universidad Carlos III de Madrid
    \AND C. Vladimir Rodríguez-Caballero~\orcidlink{0000-0002-3601-7715}\\Department of Statistics, ITAM\\Department of Economics, Duke University
     \And Esther Ruiz~\orcidlink{0000-0002-5944-9449}\\Universidad Carlos III de Madrid}
\Plainauthor{Gian Pietro Bellocca, Ignacio Garrón, Vladimir Rodríguez-Caballero, Esther Ruiz}

\title{\pkg{FARS}: Factor Augmented Regression Scenarios in \proglang{R}}
\Plaintitle{fars: Factor Augmented Regression Scenarios in R}
\Shorttitle{Factor Augmented Regression Scenarios in \proglang{R}}

\Abstract{
 In the context of macroeconomic/financial time series, the \pkg{FARS} package provides a framework in \proglang{R} for the construction of conditional densities of the variable of interest based on the factor-augmented quantile regressions (FA-QRs) methodology. Within this context, the factors used to estimate the quantiles are extracted from a multi-level dynamic factor model with potential overlapping group-specific factors, while the densities are obtained by matching the estimated quantiles to a Skewed-Student density. The package also allows the construction of measures of risk as well as designing economic scenarios for the conditional densities. In particular, the package enables users to: (i) extract global and group-specific factors using a flexible multi-level factor structure and compute asymptotically valid confidence regions for the estimated factors, accounting for uncertainty in the factor loadings; (ii) obtain estimates of the parameters of the FA-QRs together with their standard deviations, and recover full predictive conditional densities from estimated quantiles; (iii) obtain risk measures based on extreme quantiles of the conditional densities; and (iv) estimate the conditional density and the corresponding extreme quantiles when the factors are stressed.
}

\Plainkeywords{multilevel factor model, stressed factors, scenario analysis, R}

\Address{
  Gian Pietro Bellocca, Ignacio Garrón, Esther Ruiz \\
  Department of Statistics\\
  Universidad Carlos III de Madrid\\
  E-mail: \email{gbellocc@est-econ.uc3m.es, igarron@est-econ.uc3m.es, ortega@est-econ.uc3m.es}\\\\
  C. Vladimir Rodríguez-Caballero \\
  Department of Statistics\\
  Instituto Tecnológico Autónomo de México\\
    Department of Economics\\
  Duke University\\
  E-mail: \email{vladimir.rodriguez@itam.mx}\\
}

\begin{document}



\section[Introduction]{Introduction} \label{sec:intro}

In the context of macroeconomic/financial time series, there is a growing interest in obtaining predictions of the probability densities of specific key variables; see, for example, \cite{granger2000a} and \cite{granger2000b}, who argue that point forecasts are not sufficient from the perspective of a properly informed decision-maker. In addition to being of interest in themselves, these densities can also serve to obtain measures of macroeconomic vulnerability, which are crucial for the design of resilience policies; see, for example, \cite{Monache2024}. Furthermore, econometricians, policy makers, and financial analysts are also interested in the construction of realistic scenarios for the distribution of key variables that can help to further understand the resilience of economic systems by providing early warning signals of what to expect should such conditions materialize in adverse outlooks; see, for example, \cite{Gonzlez-Rivera2024} and \cite{Adrian2025}. 

To start with, estimation of the conditional density of interest is often based on assuming that underlying economic and/or financial latent factors drive it. As proposed by \cite{BaiandNg2008} and \cite{Ando2011QuantileCriterion}, and popularized by \cite{Adrian2019}, the quantiles of the distribution of the target variable can be estimated by fitting factor-augmented quantile regressions (FA-QRs) with underlying latent factors, which summarize economic and/or financial activity, as regressors. The FA-QR model allows for different impacts of underlying factors on different quantiles of the distribution of the variable of interest, and consequently, for potential asymmetries in the downside and upside risks. After estimating the quantiles, and following \cite{Azzalini2003}, the corresponding $h$-step-ahead conditional density is obtained by fitting a skew-t distribution to them. The skew-t distribution has been shown to be flexible enough to provide an appropriate approximation to the conditional density of a large number of economic variables; see \cite{mitchell2024} for alternative estimators of densities, which are shown to outperform the popular skew-t distribution in the unlikely case of multimodal distributions. The estimated conditional density delivers any quantile of interest, and, in particular, extreme quantiles, which are often used as measures of vulnerability as, for example, growth at risk (GaR) proposed by \cite{Adrian2019}, or inflation at risk (IaR) as in \cite{lopez2024}.

The factors needed as regressors for FA-QRs can be extracted from a dynamic factor model (DFM), with the preferred estimation method being principal components (PC); see, for example, \cite{Bai2003InferentialDimensions} and \cite{Bai2013} for technical details.\footnote{Note that on top of being used as predictors of FA-QRs, there are many other applications in which the factors can be of interest in themselves as, for example, when using them to construct economic/financial indexes or as predictors of diffusion indexes; see the survey on DFMs by \cite{stock2011}.} Over the last few decades, when dealing with large systems of economic variables, it is not unusual to empirically observe that some of the latent factors, which summarize the common movements in the system, only load on particular groups of variables. This block structure may represent economic, geographical, cultural, or other characteristics. In this context, PC may face difficulties. Alternatively, factors can be extracted from multi-level dynamic factor models (MLDFMs) with the matrix of factor loadings subjected to the adequate blocks of zero restrictions. The factor structure of the MLDFM allows for pervasive (or global) factors that are common across all variables in the system, as well as group-specific factors associated with one or more blocks of variables. The MLDFM can incorporate non-overlapping or overlapping blocks of variables. The factors of MLDFMs can be extracted using the sequential least squares (SLS) estimator proposed by \cite{Breitung2016} for non-overlapping factors and generalized by \cite{Rodrguez-Caballero2019} to overlapping factors. It is also important to note that, when the extracted factors are used as regressors of predictive regressions, obtaining measures of their uncertainty becomes relevant; see, for example, \cite{Amburgey2022} and \cite{Lewis2022}. The asymptotic distribution of the factors extracted by SLS is established by \cite{Choi2018} for DFMs without overlapping factors and by \cite{Lu2025} for overlapping factors.

Finally, in order to generate stressed scenarios (or stressed factors) for the conditional densities, the methodology proposed by \cite{Gonzlez-Rivera2019} can be used. Under unexpected and rare circumstances, the factors driving the distribution of the variable of interest are under stress, and thus deviate substantially from their averages. Stressed factors are probabilistically derived based on their multidimensional distribution, focusing on the observations located on its extreme autocontours.

This paper presents the \pkg{FARS} package, which provides a framework in \proglang{R} for modeling and forecasting conditional densities based on MLDFMs and FA-QRs.\footnote{Version~0.8.0  of the \pkg{FARS} package is available in CRAN: \url{https://CRAN.R-project.org/package=FARS}.} The package enables users to:
\begin{enumerate}
\item Use SLS to extract pervasive, semipervasive, and block-specific factors based on a flexible specification of the MLDFM. This approach allows for overlapping blocks, enabling the extraction of factors within a hierarchical structure with more than two levels, beyond the usual global-local configuration. To the best of our knowledge, there are no other published \proglang{R} packages that provide this functionality. The only closely related package available in \proglang{R} is \pkg{GCCfactor} by \cite{GCCfactor}, which allows factor estimation in MLDFMs with non-overlapping structures. The estimation method is based on the generalized canonical correlation (GCC) estimator, which is conceptually related to SLS; see \cite{Lin2022}.\footnote{Other implementations of DFMs (not multilevel) are available in the \proglang{R}, although they are not published. The \pkg{sparseDFM} package implements popular estimation methods for DFMs, including the recent Sparse DFM approach by \cite{Mosley2024}; see \cite{sparseDFM}. The \pkg{MARSS, KFAS} packages provide a flexible framework for modeling DFMs within state-space structures (\cite{MARSS} and \cite{helske2017kfas}). Furthermore, the \pkg{dfms} package offers a broad suite of DFM estimation techniques under the assumption of idiosyncratic components independently and identically distributed (\textit{i.i.d.}) \citep{dfms}. Also, commercial software can be used to extract factors from DFMs in \proglang{Eviews}, \proglang{MATLAB}, \proglang{Stata} and \proglang{Julia}; see, for example, \cite{solberger2020} for the estimation of the DFM in the context of state-space models.} Alternatively, several open-source packages provide functionalities for multilevel modeling using different methods, including \pkg{lavaan} \citep{rosseel2012lavaan} and \pkg{OpenMx} \citep{Boker_2021} in \proglang{R}, as well as the \pkg{mfmodel} package in \proglang{Python} \citep{parshakova2025fitting}. Commercial implementations are also available, such as \pkg{GLLAMM} in \proglang{Stata} and in \proglang{MLwiN}; see \cite{rabe2004gllamm} and \cite{rasbash2000user}, respectively. 

\item Compute asymptotically valid confidence regions for the SLS factors, accounting for uncertainty in the factor loadings, and for potential cross-correlations of the idiosyncratic components. As far as we are concerned, the only software that allows inference on factors is \pkg{GCCfactor}, where the confidence regions are based on bootstrap instead of being asymptotic.

\item Estimate FA-QR models and recover full predictive conditional densities from the estimated quantiles using the quantile-matching method to determine the parameters of a skew-t distribution. Finally, obtain risk measures such as GaR and IaR. A similar implementation is presented by \cite{lajaunie2025}, which provides an unpublished software in \proglang{R} to estimate factor-augmented quantile regressions and the corresponding densities.\footnote{There are alternative \proglang{R} packages for estimating quantile regressions (including linear, nonlinear, censored, locally polynomial, and additive models) such as the \pkg{quantreg} package by \citet{quantreg}. However, these implementations do not include factor-augmented specifications. Conversely, packages like \pkg{pls} by \citet{mevik2007} allow for factor-augmented regressions but do not estimate quantiles.} Furthermore, the International Monetary Fund offers a \proglang{Microsoft Excel} toolkit that integrates macros and a user interface to perform GaR estimation, based on \cite{Adrian2019}; see \cite{Prasad2019} for further details. This toolkit relies on a \proglang{Python} backend that performs estimation, optimization, distribution fitting, simulations, and visualization.

\item Obtain scenarios for the conditional density and associated risk measures when the factors are stressed (via extreme autocontours of their joint distribution). None of the alternative software mentioned above implements such functionality.

\end{enumerate}

The main computational challenge addressed in this paper is the flexibility of the factor extraction procedure to accommodate a wide range of model specifications with potentially overlapping factor structures. Previously, the code used to extract the factors had to be adapted for each particular model specification. This made the implementation of the procedure complicated for interested users, who needed to modify their code whenever the factor structure changed. Second, although also an important advantage, the design of \pkg{FARS} is modular, allowing users to supply their own factors or estimated densities and integrate them to obtain the remaining outputs. Third, the package implements subsampling by blocks, which introduces additional computational challenges. Finally, packaging the entire workflow in an open-source R package substantially lowers the barrier for applied researchers and policy institutions to implement advanced density-based methods.

The functionalities of the \pkg{FARS} package are illustrated through two different examples. In the first example, the underlying factors of headline inflation are extracted across a large number of European countries. The factors are subsequently used to estimate the conditional density of aggregate inflation in a selected country to assess the associated risk of high inflation, both when the economy is under business-as-usual conditions and when it is under stress. A second example considers building scenarios for the density of economic growth in the United States (US), as in \cite{Gonzlez-Rivera2024}.

The rest of this paper is organized as follows. The methodology is briefly described in Section~\ref{sec:methodology}. Section~\ref{sec:FARS_package} describes the functionality of the \pkg{FARS} package. Section~\ref{sec:illustration} is devoted to illustrating two empirical applications regarding European inflation and US GDP growth. Finally, Section~\ref{sec:summary} concludes.



\section{Methodology}
\label{sec:methodology}

In this section, we provide a brief description of the methodology for extracting underlying factors and obtaining conditional density forecasts of the target variable under standard economic dynamics and stressed scenarios of the underlying factors. First, we discuss the factor structures involved in the DFMs and MLDFMs (with and without overlapping blocks), and describe the asymptotic distribution of the PC estimated factors, assuming that idiosyncratic components are either cross-sectionally uncorrelated or weakly correlated. Second, we describe how to obtain forecasts of the density of the target variable under both stressed and non-stressed scenarios using FA-QRs.

\subsection{Dynamic factor model (DFM)}

The DFM has been extensively studied in the literature to reduce the dimensionality of large sets of variables by assuming that they can be represented by a relatively small number of common underlying factors; see, for example, \cite{Stock2002b,Stock2002a}, \cite{Bai2003InferentialDimensions}, and \cite{Bai2013}. Consider \( \mathbf{X}_t=(x_{1t},...,x_{Nt})^\top \), the \( N \times 1 \) vector of weakly stationary variables observed at time \( t=1,...,T \). The DFM is given by
\begin{equation}
\mathbf{X}_t = \mathbf{P F}_t + \boldsymbol{\varepsilon}_t,
\label{eq:DFM_equation}
\end{equation}
where \( \mathbf{P}=(\mathbf{p}_1^\top,...,\mathbf{p}_N^\top)^\top\) is the \( N \times r \) matrix of factor loadings, \( \mathbf{F}_t = (F_{1t}, \dots, F_{rt})^\top \) is an \( r \times 1 \) vector of weakly stationary latent factors, and \( \boldsymbol{\epsilon}_t=(\epsilon_{1t},...,\epsilon_{Nt})^\top\) is the \( N \times 1 \) vector of idiosyncratic components, which are assumed to be weakly stationary and cross-sectionally weakly correlated, and uncorrelated with the common factors \( \mathbf{F}_t \). Finally, the number of factors, $r$, is known. 

In the model reported in Equation~\ref{eq:DFM_equation}, the loadings and factors cannot be separately identified. They can only be estimated consistently up to a rotation of the factor space. Consequently, the standard identification restrictions often assumed in the literature are that $\frac{1}{T}\mathbf{F}^\top\mathbf{F} = \mathbf{I}_r$, and that $\frac{1}{N}\mathbf{P}^\top\mathbf{P}$ is a diagonal matrix with distinct elements on the main diagonal, ordered from largest to smallest. Under these restrictions, estimated factors are identified up to a sign transformation; see \cite{Bai2013} for more details about the identification of DFMs in the context of PC estimation. 

In practice, factors are often estimated using PC. Let \( \mathbf{X} = (\mathbf{X}_1, \dots, \mathbf{X}_T)^\top \) denote the \( T \times N \) matrix of observed data. The PC-estimated factors, \( \hat{\mathbf{F}}_t \), are obtained as \( \sqrt{T} \) times the eigenvectors associated with the \( r \) highest eigenvalues of the matrix \( \mathbf{XX}^\top \), ordered in decreasing magnitude. The corresponding loading matrix is then estimated by \(\hat{\mathbf{P}}^\top = \frac{1}{T} \hat{\mathbf{F}}^\top \mathbf{X}.\)

\subsection{Multi-level dynamic factor model (MLDFM)}

In many economic/financial applications, the variables in $\mathbf{X}_t$ are naturally grouped into blocks, such as countries, geographical regions, or economic sectors. In some cases, not all variables in $\mathbf{X}_t$ load onto all factors in the DFM, which implies the presence of zeros in the matrix of loadings, $\mathbf{P}$. The standard PC approach is suboptimal in this context, as it neglects the block structure. Consequently, when the block structure is known, a more appropriate approach is to extract the factors from a MLDFM, where the relevant zero restrictions are imposed directly on $\mathbf{P}$. In what follows, we present two alternative specifications of the MLDFM, depending on whether the blocks of variables overlap.\footnote{It is important to note that the \pkg{FARS} package assumes that the MLDFM structure is known in the sense that the block structure and the number of global factors and factors within each block are previously determined by the user.}

\subsubsection{MLDFM without overlapping blocks}
 \cite{Breitung2016} propose the following MLDFM with non-overlapping blocks
 
\begin{equation}
\label{eq:multilevel_def}
\begin{bmatrix}
\mathbf{X}_{1t} \\
\vdots \\
\mathbf{X}_{Kt}
\end{bmatrix} =\begin{bmatrix}
\boldsymbol\mu_{1} & \boldsymbol\lambda_1 & 0 & \dots & 0 \\
\boldsymbol\mu_{2} & 0 & \boldsymbol\lambda_2 & \dots & 0 \\
\vdots &  & 0 & \ddots & 0 \\
\boldsymbol\mu_{K} & 0 & 0 & \dots & \boldsymbol\lambda_{K} \\
\end{bmatrix} \begin{bmatrix}
\mathbf{G}_{t}\\
\mathbf{F}_{1t}\\
\mathbf{F}_{2t}\\
\vdots \\
\mathbf{F}_{Kt}\\
\end{bmatrix} + \begin{bmatrix}
\boldsymbol{\varepsilon}_{1\cdot t}\\
\vdots \\
\boldsymbol{\varepsilon}_{K\cdot t}\\
\end{bmatrix}, \\
\end{equation}
where, for $k=1,...,K$, $\mathbf{X}_{kt}$ is the $ N_k\times1$ vector of variables within block $k$, such that the cross-sectional dimension of $\mathbf{X}_t=\left(\mathbf{X}_{1t},...,\mathbf{X}_{Kt} \right)^{\prime}$ is $N=\sum_{k=1}^K N_k$, and $\boldsymbol{\mu}_i$ and $\mathbf{\lambda}_i$ are $N_i \times r_g$ and $N_i \times r_k$ loading matrices of the global and block-specific factors on the variables within block $i$, respectively. Furthermore, \( \mathbf{G}_t = (\mathbf{G}_{1t}, \dots, \mathbf{G}_{r_Gt})^\top\) is the $r_G \times 1$ vector of pervasive factors that load on all variables in the system, while \( \mathbf{F}_{kt} = (\mathbf{F}_{1t}, \dots,  \mathbf{F}_{r_kt})^\top\) is the $r_k \times 1$ vector of block-specific factors that load only within the block $\mathbf{X}_{kt}$. The idiosyncratic noise are defined conformably; see \cite{wang2010}, \cite{Breitung2016}, \cite{Choi2018} and \cite{bellocca2026}, for further technical details and identification conditions.

\subsubsection{MLDFM with overlapping blocks}

\begin{figure}[!h]
\centering
\includegraphics[width=0.3\textwidth]{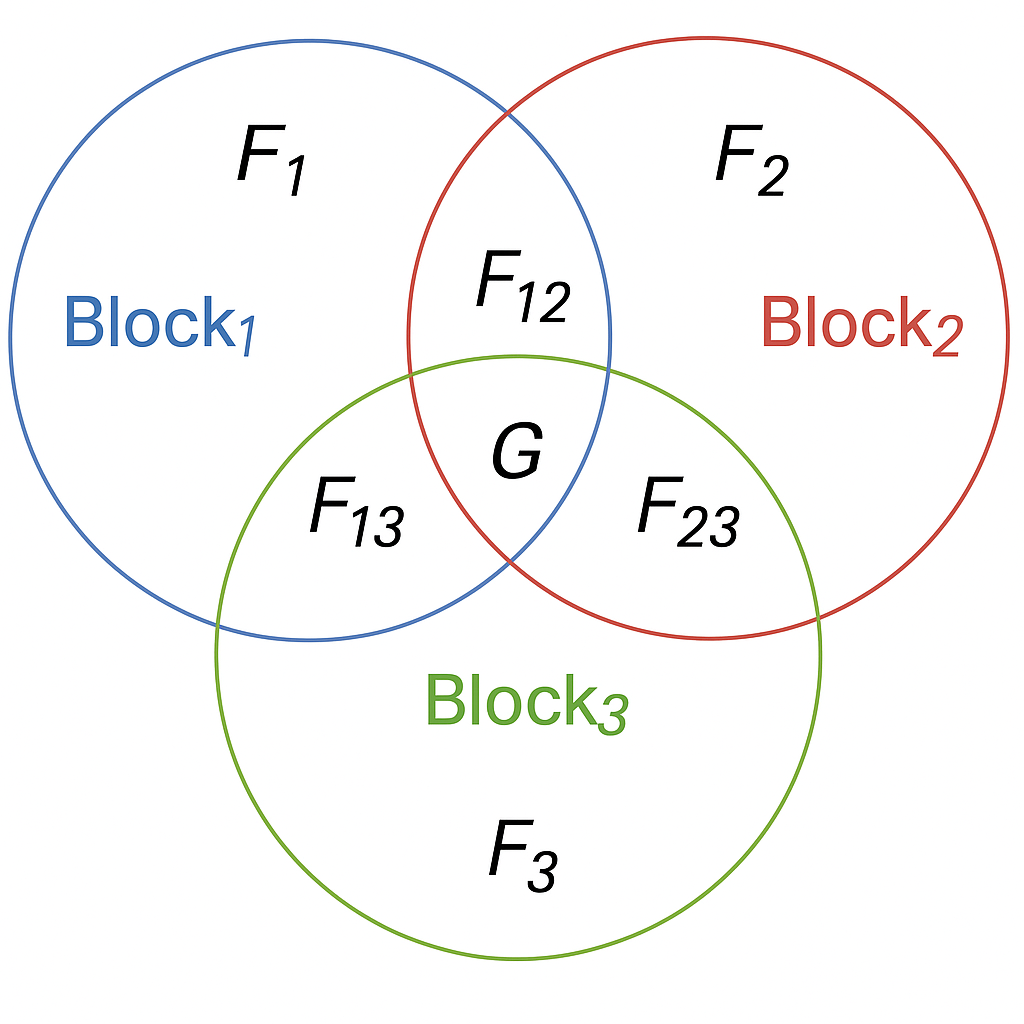}
\caption{Factor structure of the MLDFM with three different overlapping blocks of data.}
\label{fig:factor_structure}
\end{figure}

For clarity of exposition of the MLDFM with overlapping blocks, consider the case with $K=3$; see \cite{Rodrguez-Caballero2019} for a detailed description.\footnote{The \pkg{FARS} package supports $K > 3$ blocks, including triple-wise (and higher-order) interactions. However, the computational burden naturally increases when the number of blocks and/or the order of interactions increases.} Assume the presence of pervasive factors, \( \mathbf{G}_t \), and block-specific factors, \( \mathbf{F}_{k,t} = \left( \mathbf{F}^{^\top}_{1t}, \mathbf{F}^{^\top}_{2t}, \mathbf{F}^{^\top}_{3t} \right)^{^\top} \), as described above. In addition, a general factor structure may also include pairwise (or semipervasive) factors, \( \mathbf{F}_{kj,t} = \left( \mathbf{F}^{^\top}_{12,t}, \mathbf{F}^{^\top}_{13,t}, \mathbf{F}^{^\top}_{23,t} \right)^{^\top} \). For instance, the factor \(\mathbf{F}_{12,t} \) loads on the variables in blocks $\mathbf{X}_{1t}$ and $\mathbf{X}_{2t}$; that is, the semipervasive factor captures the commonality only between blocks $1$ and $2$ without any dependence on block $3$. This type of factor structure is illustrated in Figure~\ref{fig:factor_structure}, which represents the relationships between pervasive, semipervasive, and block-specific factors, when $K=3$. 

The MLDFM with overlapping blocks is given by
\begin{equation*}
x_{kit} = \boldsymbol{\mu}^\top_{ki}\mathbf{G}_{t} + \boldsymbol{\kappa}_{k j_{i}}^\top\mathbf{F}_{k j,t}+\boldsymbol{\lambda}_{ki}^\top\mathbf{F}_{kt}+ \varepsilon_{kit},
\end{equation*}
where $k=1,2,3$ indicates the block, index $i=1,\ldots, N_k$ denotes the $i'th$ cross-section unit of block $k$, $t=1,\ldots , T$ is the period of time, and $kj$ means interaction between blocks $k$ and $j$ $\in (1,2,3)$ with $k\neq j$. $\boldsymbol{\mu}_{ki}, \boldsymbol{\kappa}_{k j_{i}},$ and $\boldsymbol{\lambda}_{ki}$ are the $r_G, r_{F_{kj}},$ and $r_{F_{k}}$- dimensional vectors of factor loadings. The number of pervasive, pairwise, and block-specific factors can naturally vary in each block $k$. The idiosyncratic term denoted by $\varepsilon_{k,it}$ satisfies the standard assumptions of the DFM defined in Equation~\ref{eq:DFM_equation}. 

The vector representation of the three-block MLDFM with overlapping blocks is given by
\begin{equation}
\begin{aligned}
\begin{bmatrix}
\mathbf{X}_{1t} \\
\mathbf{X}_{2t} \\
\mathbf{X}_{3t}
\end{bmatrix}
&=
\begin{bmatrix}
\boldsymbol\mu_1 & \boldsymbol\kappa_{12_1} & \boldsymbol\kappa_{13_1} & 0 & \boldsymbol\lambda_1 & 0 & 0 \\
\boldsymbol\mu_2 & \boldsymbol\kappa_{12_2} & 0 & \boldsymbol\kappa_{23_2} & 0 & \boldsymbol\lambda_2 & 0 \\
\boldsymbol\mu_3 & 0 & \boldsymbol\kappa_{13_3} & \boldsymbol\kappa_{23_3} & 0 & 0 & \boldsymbol\lambda_3
\end{bmatrix}
\begin{bmatrix}
\mathbf{G}_t \\
\mathbf{F}_{12,t} \\
\mathbf{F}_{13,t} \\
\mathbf{F}_{23,t} \\
\mathbf{F}_{1t} \\
\mathbf{F}_{2t} \\
\mathbf{F}_{3t}
\end{bmatrix}
+
\begin{bmatrix}
\boldsymbol{\varepsilon}_{1t} \\
\boldsymbol{\varepsilon}_{2t} \\
\boldsymbol{\varepsilon}_{3t}
\end{bmatrix}.
\end{aligned}
\label{eq:ML_over}
\end{equation}
Note that the total number of unobserved common factors involved in Equation~\ref{eq:ML_over} is $r_G+r_{F_{12}}+r_{F_{13}}+r_{F_{23}}+r_{F_{1}}+r_{F_{2}}+r_{F_{3}}$. \cite{hallin2011dynamic} and \cite{ergemen2023estimation} propose a simple methodology based on the inclusion-exclusion principle of set theory to determine the number of pervasive, semipervasive and block-specific factors. However, the \pkg{FARS} package assumes that this number is known.

\subsubsection{Sequential least squares estimation}

The estimation of the MLDFM is based on the SLS approach proposed by \cite{Breitung2016} in which the main goal is to minimize the following residual sums of squares (RSS) function:

\begin{equation}
S(\hat{\mathbf{F}_t}, \hat{\mathbf{P}}) = \sum_{t=1}^{T} \left( \mathbf{X}_t - \hat{\mathbf{P}} \hat{\mathbf{F}_t} \right)^\top \left( \mathbf{X}_t - \hat{\mathbf{P}} \hat{\mathbf{F}_t} \right),
\end{equation}
by a sequence of LS regressions. The algorithm can be executed for the general case of $K$ blocks with overlapping factors as follows:

\begin{enumerate}
    \item Obtain initial values of the factors as follows:
    \begin{enumerate}
        \item Employ canonical correlation analysis (CCA) on $\mathbf{X}_{k,t}$ to obtain initial estimates of the global factor, $\hat{\mathbf{G}}^{(0)} = \left( \hat{\mathbf{G}}^{(0)}_1, \hat{\mathbf{G}}^{(0)}_2, \dots, \hat{\mathbf{G}}^{(0)}_T \right)^\top$.
        \item Filter out the global component by regressing $\mathbf{X}_{k,t}$ on $\hat{\mathbf{G}}^{(0)}$, and get the corresponding residuals, $\mathbf{X}^{*(0)}_{kt}$, from each of the $K$ separate regressions.
        \item Employ CCA on $\mathbf{X}^{*(0)}_{kt}$ to obtain the following lower-level factors, selecting the corresponding blocks.
        \item Regress $\mathbf{X}^{*(0)}_{kt}$ on the respective lower-level factors involved and get the residuals.
        \item Steps c) and d) are executed sequentially until the initial estimates of the pairwise block factors are obtained. Denote by $\mathbf{X}^{**(0)}_{kit}$ the residuals after filtering the pairwise factors of each block $k$.
        \item Run PC on $\mathbf{X}^{**(0)}_{k,t}$ to obtain the block-specific factors $\hat{\mathbf{F}}^{(0)} = \left( \hat{\mathbf{F}}^{(0)}_{1t}, \hat{\mathbf{F}}^{(0)}_{2t}, \dots, \hat{\mathbf{F}}^{(0)}_{kt} \right)^\top$.
        \item The initial loading matrix, $\hat{\mathbf{P}}^{(0)}$, is estimated through time-series regressions of $\mathbf{X}_{k,t}$ on global factors, $\mathbf{X}^*_{kt}$ on semi-pervasive factors, and $X^{**}_{kt}$ on non-pervasive factors.
    \end{enumerate}
    \item Updated estimates for the unobserved factors, $\hat{\mathbf{F}}^{(1)}$, are obtained by LS regression of $\mathbf{X}_{kt}$ on $\hat{\mathbf{P}}^{(0)}$ as follows $\hat{\mathbf{F}}^{(1)} = \left( \hat{\mathbf{P}}^{(0)^\top}\hat{\mathbf{P}}^{(0)} \right) ^{-1} \hat{\mathbf{P}}^{(0)^\top}\mathbf{X}_{k,t}$.
    \item The updated factors $\hat{\mathbf{F}}^{(1)}$ are used to obtain the associated loadings matrix, $\hat{\mathbf{P}}^{(1)}$, as in Step 1.
    \item Steps 2 and 3 are repeated until the RSS converges to a minimum, from which $\hat{\mathbf{F}}^{*}$ and $\hat{\mathbf{P}}^{*}$ are obtained.
\end{enumerate}

Note that the SLS final estimates of the pervasive, semipervasive and non-pervasive factors estimate the space spanned by the factors and not the factors themselves. Therefore, particular rotations of interest of the factors may also be obtained after estimation. The \pkg{FARS} package provide the rotation for which the usual identification conditions are satisfied.

In step 1 of the algorithm, the initialization of $\mathbf{P}^{*}$ and $\mathbf{F}_t^{*}$ is carried out using CCA. Alternatively, the \pkg{FARS} package provides the alternative of using PC. Although both approaches produce approximately the same estimated common components $\hat{\mathbf{P}}^{*} \hat{\mathbf{F}}_t^{*}$, the convergence of CCA is typically faster, requiring fewer iterations to minimize RSS. However, when the factor structure is highly complex, initializing with PC tends to be computationally more efficient; see also \cite{Breitung2016} for the comparison of the small sample properties of the SLS estimator initialized with CCA and PC for the two-level DFM.

\subsection{Asymptotic distribution of factors}

The construction of probabilistic scenarios for the unobserved factors requires knowledge of their joint distribution. The asymptotic distribution of PC factors obtained from the DFM in Equation~\ref{eq:DFM_equation} is derived by \cite{Bai2003InferentialDimensions}. If $\frac{\mathbf{F}^{\prime}\mathbf{F}}{T}=\mathbf{I}_r$ and $\frac{\sqrt{N}}{T}\rightarrow 0$ when $N, T \rightarrow \infty$, the asymptotic distribution of $\widehat{\mathbf{F}}_t$, at each moment, $t$, is given by\footnote{Note that PC factors satisfy the identification restrictions. As such, they estimate the space spanned by the true factors and not the true factors themselves. Then, it is always possible rotating the factors to obtain an interpretation of interest. In this case, the asymptotic distribution should be accordingly modified.} 

\begin{equation}
\label{eq:Asympt}
\sqrt{N} \left(\widehat{\mathbf{F}}_t - \mathbf{F}_t \right)  \overset{d}{\rightarrow } N\left( 0,\mathbf{\Sigma}_{P}^{-1}\mathbf{\Gamma} _{t}\mathbf{\Sigma}_{P}^{-1}\right), 
\end{equation}
where $\mathbf{\Sigma}_P=\lim_{N \rightarrow \infty} \frac{\mathbf{P}^{\prime}\mathbf{P}}{N}$ and $\mathbf{\Gamma}_t=\lim_{N\rightarrow \infty} \sum_ {i=1}^N \sum_{j=1}^N \mathbf{p}_i \mathbf{p}_j^{\prime} E(\varepsilon_{it} \varepsilon_{jt})$ with $\mathbf{p}_i$ and $\varepsilon_{it}$ defined as in the DFM in Equation~\ref{eq:DFM_equation}. The finite sample approximation of the asymptotic covariance matrix of $\widehat{F}_t$ can be estimated as follows:
\begin{equation}
\label{eq:MSE}
MSE_t=\left(\frac{\hat{\mathbf{P}}^{\prime}\hat{\mathbf{P}}}{N} \right)^{-1}\frac{\hat{\mathbf{\Gamma}}_t}{N} \left(\frac{\hat{\mathbf{P}}^{\prime}\hat{\mathbf{P}}}{N} \right) ^{-1},
\end{equation}
where $\hat{\mathbf{\Gamma}}_t$ is a consistent estimator for $\mathbf{\Gamma}_t$. Under the assumption of cross-sectionally uncorrelated idiosyncratic components, \cite{BaiNg2006} propose the following estimator:
\begin{equation} \label{eq:gamma}
\widehat{\mathbf{\Gamma} }^{BN}_{t}=\frac{1}{N}\sum_{i=1}^{N}\hat{\mathbf{p}}_{i}\hat{\mathbf{p}}_{i}^{^{\prime }}\hat{\varepsilon}_{it}^{2},
\end{equation}
where $\hat{\varepsilon}_{it}=x_{it}-\hat{\mathbf{p}}_{i}^{^{\prime}}\hat{\mathbf{F}_{t}}$ are the residuals from the DFM model.

In many empirical settings, assuming that the idiosyncratic covariance matrix $\Sigma_\epsilon$ is diagonal imposes a stringent restriction that may not hold in practice. Therefore, alternatively, we relax this assumption allowing the idiosyncratic components to be weakly cross-sectionally correlated. Under these circumstances, $\Gamma_t$ can be consistently estimated as proposed by \cite{Fresoli2024} using adaptive thresholding of the sample covariances of the idiosyncratic residuals, $\hat{\sigma}_{ij}$, as follows:
\begin{equation}
\label{eq:Gamma_new}
\widetilde{\mathbf{\Gamma}}^{FPR}= \frac{1}{N} \sum_{i=1}^N \sum_{j=1}^N \hat{\mathbf{p}}_i \hat{\mathbf{p}}^{\prime}_j \frac{1}{T} \sum_{t=1}^T \hat{\varepsilon}_{it} \hat{\varepsilon}_{jt} I\left(\mid \hat{\sigma}_{ij}\mid \geq c_{ij} \right),
\end{equation}
where $I(\cdot)$ is the indicator function that takes value one when the argument is true and zero otherwise, and $c_{ij}=\delta \omega_{NT} \left[ \widehat{Var}\left[\hat{\varepsilon}_{it} \hat{\varepsilon}_{jt} \right] \right]^{1/2}$, with $\widehat{Var}\left[\hat{\varepsilon}_{it} \hat{\varepsilon}_{jt} \right]= \frac{1}{T}\sum_{t=1}^T \left[\hat{\varepsilon}_{it} \hat{\varepsilon}_{jt} - \hat{\sigma}_{ij} \right] ^2$, $\omega_{NT} = \frac{1}{\sqrt{N}}+\sqrt{\frac{\text{log}(N)}{T}}$, and $\delta$ chosen as proposed by \cite{qiu2019}. It is important to note that the estimator of $\mathbf{\Gamma}_t$ in (\ref{eq:Gamma_new}) requires stationarity and, consequently, is constant over time. 

Regardless of whether $\mathbf{\Gamma}_t$ is obtained from (\ref{eq:gamma}) or (\ref{eq:Gamma_new}), the estimated asymptotic covariance matrix in (\ref{eq:MSE}) does not account for the uncertainty arising from the estimation of the loading matrix. In this light, \cite{Maldonado2021} propose a correction of the asymptotic MSE based on subsampling subsets of series of size $N^*<N$ in the cross-sectional space, with each series in the subsample containing all temporal observations. For each subsample, the loadings and factors are estimated by PC, obtaining $\hat{\mathbf{F}}_t^{*(s)}$ and $\hat{\mathbf{P}}^{*(s)}$, for $s=1,...,S$. The corrected finite sample approximation of the asymptotic MSE of $\hat{\mathbf{F}}_t$ can be estimated as follows:
\begin{equation} \label{eq:Subsampling_MSE_2}
MSE^{\ast}_t =\frac{1}{N}\left(\frac{\hat{\mathbf{P}}^{\prime}\hat{\mathbf{P}}}{N} \right)^{-1}  \hat{\mathbf{\Gamma}}_t \left(\frac{\hat{\mathbf{P}}^{\prime}\hat{\mathbf{P}}}{N} \right)^{-1} + \frac{N^*}{NS}\sum_{s=1}^{S}\left( \left( \hat{\mathbf{F}}_{t}^{\ast (s)}-\hat{F}_{t}\right) \left( \hat{\mathbf{F}}_{t}^{\ast (s)}-\hat{\mathbf{F}}_{t} \right)^{\prime }\right).
\end{equation}

Based on the asymptotic normality in (\ref{eq:Asympt}), \cite{Maldonado2021} construct confidence ellipsoids for the estimated factors with coverage probability $100\times \alpha\%$ as follows:

\begin{equation} \label{eq:ellipsoid}
g(\mathbf{F}_t,\alpha)=\{\mathbf{F}_t\in{\rm I\!R}^r |(\mathbf{F}_t-\hat{\mathbf{F}}_t) MSE_t^{*-1}(\mathbf{F}_t-\hat{\mathbf{F}}_t)\leq \chi^2_{r(\alpha)}\},
\end{equation}
where $\chi^2_{r(\alpha)}$ is the $\alpha$-quantile of the $\chi^2$ distribution with $r$ degrees of freedom, with $r$ being the number of factors. Each point on the surface of the ellipsoid represents a possible joint realization of all factors in the DFM. These boundary points correspond to extreme, yet plausible, stress conditions.





\subsection{Density forecasts under stressed and non-stressed conditions}

The estimated factors, which summarize the information contained in a large set of predictors $X_t$, are used to estimate the temporal evolution of the conditional density of a target variable. In this section, we describe how these densities can be obtained under stressed and non-stressed conditions for the underlying factors.

Let $y_t$ be the observation at time $t$ of the target variable. We start by obtaining $h$-step-ahead forecasts of the $\tau^*$-quantile of the conditional distribution of $y_t$ by estimating the following FA-QR:
\begin{equation}
q_{\tau^*}(y_{t+h} \mid y_t, \mathbf{F}_t) = \mu(\tau^*, h) + \phi(\tau^*, h) y_t + \sum_{k=1}^{r} \beta_k(\tau^*, h) F_{kt},
\label{eq:faqr}
\end{equation}
where $\mu(\tau^*, h)$, $\phi(\tau^*, h)$, and $\beta_k(\tau^*, h)$, for $k = 1, \ldots, r$, are parameters, and ${\mathbf{F}}_t$ is the $r \times 1$ vector of the underlying unobserved factors at time $t$. In practice, the underlying factors in (\ref{eq:faqr}) are replaced by their estimates, $\hat{\mathbf{F}}_t$, obtained as described above.

The parameters of the FA-QR model in (\ref{eq:faqr}) are estimated using the algorithm by \cite{KoenkerOrey1987}, which implements the quantile regression method originally developed by \cite{Koenker1978}. When the error terms are assumed to be independently distributed according to a Laplace distribution, the estimator coincides with the Maximum Likelihood (ML) estimator; see \cite{Ando2011QuantileCriterion}. \cite{BaiandNg2008} establishes its asymptotic normality.


The FA-QR provides estimates of the quantile function of the target variable, $\widehat{q}_{\tau^*}(y_{t+h}| y_t, \mathbf{F}_t)$, for several values of $\tau^{*}$. However, in practice, it is challenging to map these estimates into a probability distribution function because of approximation errors and estimation noise. Consequently, as in \cite{Adrian2019}, we use the skew-t distribution proposed by \cite{Azzalini2003} to smooth the quantile function and estimate the conditional density of $y_t$. The skew-t density depends on four parameters as follows:

\begin{equation}
f(y ; \mu, \sigma, \alpha, v)=\frac{2}{\sigma} st\left(\frac{y-\mu}{\sigma} ; v\right)sT\left(\alpha \frac{y-\mu}{\sigma} \sqrt{\frac{v+1}{v+\left(\frac{y-\mu}{\sigma}\right)^2}} ; v+1\right),
\end{equation}
where $st(\cdot)$ and $sT(\cdot)$ denote the probability density function and the cumulative distribution function of the Student's t distribution, respectively. The skew-t distribution is specified by its location $\mu$, scale $\sigma$, shape $\alpha$, and tail heaviness $v$. At each time $t$, a skew-t distribution is fitted by choosing the parameters that minimize the squared differences between the quantile estimates and the skew-t implied quantiles, $q_{\tau^{*}}(y ; \mu, \sigma, \alpha, v)$, as follows:

\begin{equation}
(\hat{\mu}_{t+h}, \hat{\sigma}_{t+h}, \hat{\alpha}_{t+h}, \hat{v}_{t+h})=\underset{\mu, \sigma, \alpha, v}{\operatorname{argmin}} \sum_{t=1}^{T-h}(\widehat{q}_{\tau^*}(y_{t+h} \mid y_t, \mathbf{F}_t)-q_{\tau^{*}}(y_t ; \mu, \sigma, \alpha, v))^{2}.
 \label{eq:minskew}
\end{equation}

The methodology described above estimates the conditional density of $y_t$ under non-stressed conditions. To construct conditional densities based on stressed scenarios, \cite{Gonzlez-Rivera2019} and \cite{Gonzlez-Rivera2024} use the confidence ellipsoids defined in Equation~\ref{eq:ellipsoid}, and determine the value of the factors in the $\alpha\%$-contour (stress level of the underlying factors) that minimize (or maximize) a given quantile ($\tau$) of the conditional distribution of the target variable. For instance, consider that we are interested in deriving a stress scenario for $\tau=0.05$, with the factors stressed at their $\alpha\%$ level, \pkg{FARS} solves the following optimization problem at each $t$

\begin{equation}
\label{eq:upper_optim}
{\min_{\mathbf{F}_t^{(S)}}}\;\hat{q}_{0.05}(y_{t+h}|y_t,\mathbf{F}_t^{(S)})
\end{equation}
\begin{equation*}
s.t. \quad g(\mathbf{F}_t^{(S)},\alpha)=0,
\end{equation*}
where $g(\mathbf{F}_t^{(S)},\alpha)=0$ is a predetermined $\alpha$-contour of the factors, that is, an ellipsoid that contains $\mathbf{F}_t$ with probability $\alpha$. 

The values of $\mathbf{F}_t^{(S)}$ on the boundary of the ellipsoid $g(\mathbf{F}_t^{(S)},\alpha)=0$ represent extreme events of the factors. After solving the optimization problem in Equation~\ref{eq:upper_optim}, these optimized values are plugged into the estimated FA-QRs. The conditional density of $y_t$ under stress is then obtained by smoothing the corresponding quantiles as described in Equation~\ref{eq:minskew}.\footnote{Note that the stressed scenarios are slightly different from that in \cite{Gonzlez-Rivera2019} and \cite{Gonzlez-Rivera2024}, who obtain stressed factors for each quantile of the distribution.}



\section{The FARS package}
\label{sec:FARS_package}

In this section, we provide a detailed overview of the \pkg{FARS} package functionalities and explain how users can implement the methodology described in Section~\ref{sec:methodology} using the available functions.

\subsection{MLDFM in FARS}

We begin by introducing the \code{mldfm()} function, which provides users with a flexible tool for extracting factors using DFM or MLDFM, with non-overlapping or overlapping blocks. In the case of a simple DFM, the function requires two arguments. The first is \code{data}, which contains the N variables from which the factors are extracted, structured as a $T\times N$ matrix. The second argument is \code{global}, which specifies the number of factors $r$ to be extracted from the data. By default, the input data are centered and scaled prior to factor extraction; this behavior can be controlled via the optional logical arguments \code{center} and \code{scale}, both set to \code{TRUE} by default.

In the case of the MLDFM without overlapping blocks, additional arguments must be provided to the \code{mldfm} function: i) the argument \code{blocks} defines the number of blocks $K$ that make up the data sample (the default is 1, corresponding to the DFM case); ii) \code{block_ind} requires a vector that indicates the indices of the end column for each block $k$. For example, if $K=3$ and $N = N_1 + N_2 + N_3$, the argument \code{block_ind} should contain $[N_1, N_1+N_2, N_1+N_2+N_3]$; iii) the argument \code{local} is a vector of integers, indicating the number of block-specific factors $r_{F_k}$ to be extracted from each block $k$; iv) \code{global} specifies the number of pervasive factors $r_G$; v) \code{method} defines the factor initialization strategy for the sequential LS estimation: 0 for the CCA (default) and 1 for PC\footnote{PC is implemented using the \code{prcomp()} function from the package \pkg{stats}.}; vi) the arguments \code{tol} and \code{max_iter} define the tolerance level and the maximum number of iterations allowed for the RSS minimization process, with default values set to $10^{-6}$ and $1000$, respectively.

In the case of the MLDFM with overlapping blocks, an additional \code{middle_layer} argument must be provided. \code{middle_layer} is a named list, where each name is a string specifying a group of overlapping blocks (e.g., $kj$ in the case of pairwise groups), and each value is the number of factors $r_{{kj}}$ to extract from that group. For example, if we want to extract one pairwise factor from blocks 1 and 3 ($r_{{13}}=1$), the list should be defined as \code{list("1-3" = 1)}.

Regardless of the particular specification of the model, the \code{mldfm()} function returns an S3 object of class \code{mldfm} as output. The object is a list containing several attributes described in Table~\ref{tab:mldfm_object}.

\begin{table}[h]
\centering
\begin{tabular}{ll}
\hline
Attribute & Description \\
\hline
\code{factors} & $T \times r$ matrix containing all the extracted factors. \\
\code{loadings} & $N \times r$ matrix of factor loadings with necessary zero restrictions. \\
\code{residuals} & $T \times N$ residual matrix from the model fit. \\
\code{fitted} & $T \times N$  matrix containing the fitted values $\hat{\mathbf{X}}$. \\
\code{method} & The initialization strategy used (CCA or PCA). \\
\code{iterations} & Number of iterations performed until convergence (0 in DFM). \\
\code{factors\_list} & A summary list indicating the number of factors extracted at each level. \\
\hline
\end{tabular}
\caption{A description of \code{mldfm} object attributes. The data stored in the factors and loadings matrices follow the hierarchical order (from global to local) described in \code{factors\_list}}.
\label{tab:mldfm_object}
\end{table}

The \code{mldfm} object has typical S3 methods: \code{print()}, \code{summary()} and \code{plot()}. The first two functions offer a brief overview of the model estimation outcome, while \code{plot()} offers pre-configured visualization tools. The call of the \code{plot} function on a \code{mldfm} object generates distinct line charts for all estimated factors, each enriched with confidence interval bands that assume cross-sectionally independent and homoskedastic idiosyncratic components. Furthermore, an optional input argument \code{dates} can be provided. \code{dates} is a vector of dates to be displayed on the x-axis, replacing the default integer time index ranging from 1 to $T$. An additional optional argument, \code{flip}, can be supplied to improve the interpretability of the plots. \code{flip} is a binary vector (with values \code{0} or \code{1}) indicating whether each estimated factor or the corresponding loadings should be sign-flipped before plotting. A value of \code{1} for a given element reverses its sign, while \code{0} leaves it unchanged. This is useful when the arbitrary sign indeterminacy of factor models leads to less interpretable visualizations. An optional argument, \code{fpr}, can be set to \code{TRUE} to estimate the asymptotic MSE of the factors using $\widetilde{\Gamma}^{\text{FPR}}$ as defined in Equation~\ref{eq:Gamma_new}. Differently, the default setup (\code{FALSE}) uses $\hat{\Gamma}_t^{\text{BN}}$ as described in Equation~\ref{eq:gamma}. Moreover, using the \code{plot()} function, it is possible to visualize the estimated loadings or residuals, specifying a \code{which} argument with values \code{"loadings"} or \code{"residuals"}. With \code{"loadings"}, a set of bar charts is generated, one for each factor. Each bar chart displays the estimated loadings along with their corresponding confidence intervals. Differently, with \code{"residuals"}, a figure depicting the correlation heatmap of the residuals is produced. In both cases, the user can provide a list of variable names using the optional \code{var_names} argument. This enables the replacement of the default indexes from VAR 1 to VAR N with the appropriate variable names. Specific attributes of the \code{mldfm} object can be accessed using appropriate functions, \code{factors()}, \code{loadings()}, \code{residuals()} and \code{fitted()}.

\subsection{Probability distribution of factors in FARS}

A two-step procedure is implemented in \pkg{FARS} to obtain the asymptotic joint probability density of the factors with the subsampling correction. 

The first step involves running a subsampling method to extract factors from subsets of $N^*$ variables, selected from the entire data sample. This is implemented using the \code{mldfm\allowbreak_subsampling()} function. The function iteratively generates \code{n_samples} subsamples of size \code{sample_size} and estimates factors using the MLDFM approach through the \code{mldfm()} function\footnote{The argument \code{n\_samples} is the number of samples, while \code{sample\_size} is the proportion of the cross-sectional dimension, $N$, which composes the subsamples (e.g., 0.9 to select 90\% of the original variables). In the case of multiple blocks, the proportion is maintained in all the blocks.}. This approach offers two main advantages. First, the arguments of \code{mldfm\allowbreak_subsampling()} are the same as those of \code{mldfm()}, plus two additional arguments to define the number and size of the subsamples. Second, the function returns an object \code{mldfm_subsample} containing a list of \code{mldfm} objects, enabling the user to apply standard methods to each of the subsample results. In addition, an optional \code{seed} argument can be provided to ensure the reproducibility of the results. A \code{mldfm_subsample} object contains the attributes listed in Table~\ref{tab:mldfm_subsample_object} and provides \code{print()}, \code{summary()} and \code{plot()} methods, as well as \code{get_mldfm_list()} and \code{get_mldfm_model()} functions to access the entire list or a specific \code{mldfm} object, respectively.

\begin{table}[h]
\centering
\begin{tabular}{ll}
\hline
Attribute & Description \\
\hline
\code{models} & A list containing the \code{n\_samples} \code{mldfm} objects. \\
\code{n\_samples} & The number of subsamples generated. \\
\code{sample\_size} & The proportion of the sample used for each subsample $\frac{N^{*}}{N}$. \\
\code{seed} & The seed used for random sampling. \\
\hline
\end{tabular}
\caption{A description of \code{mldfm\_subsample} object attributes.}
\label{tab:mldfm_subsample_object}
\end{table}

The second step involves constructing confidence regions for the factors, as described in Equation~\ref{eq:ellipsoid}. This operation is performed by the \code{create_scenario()} function, which requires three main arguments. The first is \code{model}, which contains the result of the \code{mldfm()} function applied to the full dataset and serves as the center of the ellipsoid. The second is \code{subsamples}, which uses the output of \code{mldfm_subsampling()} to compute the MSE correction as defined in Equation~\ref{eq:Subsampling_MSE_2}. The third is \code{alpha}, which defines the coverage probability (i.e., the level of stress) of the ellipsoids. An optional argument, \code{fpr}, can be set to \code{TRUE} to estimate the asymptotic MSE of the factors using $\widetilde{\Gamma}^{\text{FPR}}$ as defined in Equation~\ref{eq:Gamma_new}. Differently, the default setup (\code{FALSE}) uses $\hat{\Gamma}_t^{\text{BN}}$ as described in Equation~\ref{eq:gamma}. The output of \code{create_scenario()} is a \code{fars\_scenario} object whose attributes are presented in Table~\ref{tab:fars_scenario_object}. A \code{fars\_scenario} object is provided with the standard S3 methos (\code{print()}, \code{summary()} and \code{plot()}) and with \code{get_ellipsoids()} and \code{get_sigma_list()} functions to access specific arguments. In particular, \code{get_ellipsoids()} returns a list of $T$ matrices of size $z \times r$ representing the ellipsoid points in $r$ dimensions at each time $t$. The number of points $z$ depends on the number of dimensions $r$. In the case of only one factor ($r = 1$), only a confidence interval is built based on the specified \code{alpha} level; for this reason, $z = 2$ (i.e., the upper and the lower bounds). In the case of two dimensions ($r = 2$), the 2-D ellipsoid is composed of $z = 300$ points and is built using the \pkg{ellipse} package; see \cite{ellipse}. Lastly, in the case of more than two dimensions ($r > 2$), the $r$-D ellipsoid is generated through the \code{hyperellipsoid()} and \code{hypercube_mesh()} functions from the \pkg{SyScSelection} package \citep{SyScSelection}. In this case, the number of points composing the ellipsoid depends on the \code{phi} parameter of the \code{hypercube_mesh()} function, which defines the scalar fineness of the mesh. In \pkg{FARS}, \code{phi} is set to 8. 

\begin{table}[h]
\centering
\begin{tabular}{ll}
\hline
Attribute & Description \\
\hline
\code{ellipsoids} & A list containing $T$ matrices of dimensions $r \times z$. \\
\code{center} & $T \times r$ matrix containing all the factors used as center for the ellipsoids. \\
\code{sigma} & A list of $T$ covariance matrices of dimensions $r \times r$. \\
\code{periods} & Number of time periods $T$. \\
\code{n\_points} & Number of points $z$ used to define each ellipsoid. \\
\code{alpha} & Confidence level for the ellipsoids. \\
\hline
\end{tabular}
\caption{A description of \code{fars\_scenario} object attributes.}
\label{tab:fars_scenario_object}
\end{table}

\subsection{Conditional density under stressed and non-stressed conditions in FARS}

In this section, we present the tools provided by \pkg{FARS} for obtaining conditional density forecasts in both the non-stressed and stressed scenarios.

The first step is to estimate the FA-QRs\footnote{FARS estimate FA-QRs using the \pkg{quantreg} package \citep{quantreg}. The standard deviations of the estimated parameters are calculated using the sandwich formula proposed by \cite{Powell1989} under the option \code{ker}, which is commonly used in practice.}. This operation is performed through the \code{compute\allowbreak_fars()} function, which estimates the parameter of the FA-QR in Equation~\ref{eq:faqr}. The function requires only three arguments to work. First, \code{dep_variable}, which contains the dependent variable $y_t$. Second, \code{factors}, which includes the factors the user wants to add to the quantile regression model.\footnote{These can be easily accessed through the \code{factors} attribute of the \code{mldfm} object obtained after estimating the MLDFM by \code{mldfm()}.} Third, \code{h}, which defines the forecast horizon (the default is $h = 1$). The function estimates the FA-QRs for a fixed set of quantiles: 0.05, 0.25, 0.50, 0.75, and 0.95, as these are later used for the skew-t density fit. Alternatively, the user can modify the extreme quantiles by setting an optional \code{edge} argument. For example, setting \code{edge = 0.01} forces the edge quantiles to 0.01 and 0.99. The default value is 0.05. The output of \code{compute_fars()} is an S3 object of type \code{fars}, which contains a set of attributes listed in Table~\ref{tab:fars_object}. 

\begin{table}[h]
\centering
\begin{tabular}{ll}
\hline
Attribute & Description \\
\hline
\code{models} &  A list of 5 \code{rq} objects, each for a specific quantile regression. \\
\code{periods} & Number of time periods $T$. \\
\code{n\_factors} & Number of factors used in the regressions. \\
\code{h} & Lag order used in the regressions. \\
\code{levels} & The list of estimated quantiles. \\
\hline
\end{tabular}
\caption{A description of \code{fars} object attributes.}
\label{tab:fars_object}
\end{table}

Like the \code{mldfm} object, the \code{fars} object has standard S3 methods. The \code{print()} function provides a brief overview of the FA-QRs. The \code{summary()} function returns a detailed summary of each \code{rq} object by iteratively calling \code{summary()} on each of them. Lastly, the \code{plot()} function generates a line charts of the fitted values. The function can display customized dates on the x-axis by setting the corresponding optional argument \code{dates}. Additional methods are provided. A specific \code{rq} object can be accessed through \code{get_rq_model()}. The functions \code{fitted()} and \code{residuals()} return, respectively, a $T \times Q$ matrix of fitted values and residuals estimated for the five quantiles ($Q=5$). The function \code{coef()} displays the estimated coefficients for all quantiles, while \code{logLik()} returns a \code{logLik} object corresponding to a user-specified quantile. Moreover, \code{predict()} allows out-of-sample estimation for the five quantiles, taking the required input data as a matrix through the \code{newdata} argument. An important feature of the \code{predict()} function is the ability to estimate stressed quantiles by providing stressed factors in the \code{newdata} matrix instead of the baseline factors. These stressed factors can be obtained using the \code{compute_stressed_factors()} function, which implements the optimization procedure described in Section~\ref{sec:methodology} (see Equation~\ref{eq:minskew}). In brief, the function estimates a FA-QR for the selected quantile level \code{qtau} and lag order \code{h}, and identifies the factor values on the contour ellipsoid that minimize or maximize the conditional quantile of the dependent variable. The resulting stressed factors can be then used in \code{predict()} to compute quantiles under stressed scenarios.

The second step to obtaining a density forecast is to estimate the density of the target variable $y_t$ by fitting a skewed-t distribution. This operation is performed via the \code{compute_density()} function, which requires a \code{quantiles} argument, containing the quantiles estimated before\footnote{If the quantiles computed with \code{compute\_fars()} have been modified via the \code{edge} argument, the density function must be informed of the correct quantiles levels. This can be done by setting the \code{levels} argument using levels obtained by calling \code{get\_quantile\_levels()} on the \code{fars} object.}. Depending on the quantiles provided, the density function returns the non-stressed or the stressed density. Additional arguments can be provided to \code{compute_density()}, including \code{est_points}, which set the number of estimation points (default is 512), \code{random_samples}, which define the number of random samples to be drawn from the estimated distribution (default is 5000) and \code{support}, which select the lower and upper bounds of the random variable support (default is \code{c(-10,10)}). For each period $t$, \code{compute_density()} initializes the skewed-t distribution by setting three parameters (location, scale, and shape) using the quantile values provided as input. The function implements two optimization procedures to fit the skew-t distribution. The default is a linear optimization using \code{optim()} from \pkg{stats}, which implements the \code{L-BFGS-B} method. The second is a non-linear optimization method that can be selected by setting the argument \code{nl = TRUE}. The non-linear method is from the \pkg{nloptr} package and is based on \code{NLOPT_LN_SBPLX} (\cite{NLopt}). In both cases, the theoretical quantiles and the probability distribution function (pdf) of the fitted skewed-t distribution are computed using \code{qst()} and \code{dst()} from \pkg{sn} (\cite{snpkg}), respectively. Finally, a \code{seed} argument can be provided to ensure the reproducibility of the results. The \code{compute_density()} function returns a \code{fars_density} object that provides the attributes listed in Table~\ref{tab:fars_density_object}.

\begin{table}[h]
\centering
\begin{tabular}{ll}
\hline
Attribute & Description \\
\hline
\code{density} & The estimated densities at time $t$. \\
\code{distribution} & The random draws from the fitted skew-t distribution at each $t$. \\
\code{optimization} & The optimization method implemented: linear or non-linear. \\
\code{eval\_points} & The sequence of evaluation points used to compute the density. \\
\hline
\end{tabular}
\caption{A description of \code{fars\_density} object attributes. Both \code{density} and \code{distribution} are provided in matrix form with one row for each time $t$.}
\label{tab:fars_density_object}
\end{table}

The \code{fars_density} object is equipped with standard S3 methods. The \code{print()} function provides a brief overview of the estimated density. The \code{summary()} function returns the mean, median, and standard deviation of the distribution at each time $t$. Finally, the \code{plot()} function generates a 3D plot of the density, with evaluation points (\code{eval_points}) on the x-axis, time indices on the y-axis, and density values on the z-axis. The function can also display custom dates on the y-axis by setting the optional argument \code{time_index}. The distribution is accessible through the \code{get_distribution()} function.

The final step in obtaining a conditional density forecast is to extract the conditional quantile from the estimated skew-t distribution. This can be performed using the function \code{quantile_risk()}. This function requires two parameters: an object of class \code{fars_density} and the quantile that must be extracted \code{qtau}. The quantile extraction is implemented via \code{quantile()} from \pkg{stats}. Depending on the \code{fars_density} object provided, either a non-stressed or a stressed density, the \code{quantile_risk()} extracts a non-stressed quantile or a stressed quantile of the target variable (e.g., in the case of GDP growth with \code{qtau = 0.05}, it extracts growth at risk or growth in stress).

\begin{figure}[!h]
    \centering
    \includegraphics[width=0.7\textwidth]{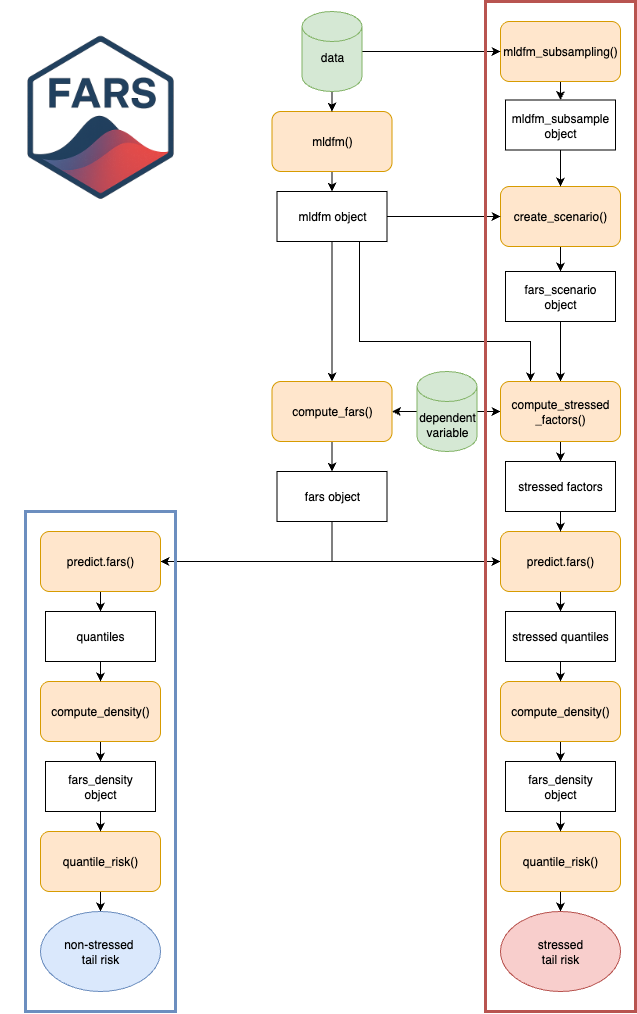}
    \caption{\pkg{FARS} package workflow for both non-stressed and stressed scenarios.}
    \label{fig:fars_flow}
\end{figure}

Figure~\ref{fig:fars_flow} shows a recap of the \pkg{FARS} package workflow for both the non-stressed and the stressed scenarios.

\section{Illustration of FARS package functionalities}
\label{sec:illustration}

In this section, we illustrate the functionalities of the \pkg{FARS} package by extracting factors, estimating conditional densities and obtaining stress scenarios in the context of: i) aggregate inflation in Europe; and ii) growth in US. Regardless of the particular application, the first step is to install and load the package \pkg{FARS}, which is publicly available on CRAN under the GPL-3 license, as follows:

\begin{CodeChunk}
\begin{CodeInput}
R> install.packages("FARS")
\end{CodeInput}
\end{CodeChunk}

The development version is available on GitHub at \url{https://github.com/GPEBellocca/FARS}. This can be downloaded using the \pkg{devtools} package with the following command:

\begin{CodeChunk}
\begin{CodeInput}
R> devtools::install_github("GPEBellocca/FARS")
\end{CodeInput}
\end{CodeChunk}

After installing the package from CRAN or GitHub, it should be loaded as follows:

\begin{CodeChunk}
\begin{CodeInput}
R> library("FARS")
\end{CodeInput}
\end{CodeChunk}

\subsection{Measuring inflation risk in Europe using FARS}
\label{sec:EA}

In the first illustration, we analyze the risk of an inflation increase in Europe. To do this, we collect monthly headline CPI data \citep{HA2023102896} from January 2005 to December 2024 ($T=240$) for a set of $N=38$ European countries. The countries considered are divided into three different blocks, depending on geographical location:
\begin{itemize}
  \item West ($N_1 = 11$): Austria, Belgium, France, Germany, Ireland, Italy, Luxembourg, Portugal, Spain, Switzerland, United Kingdom;
  \item East ($N_2 = 21$): Belarus, Bosnia and Herzegovina, Bulgaria, Croatia, Cyprus, Czech Republic, Estonia, Greece, Hungary, Kosovo, Latvia, Lithuania, Malta, Moldova, Rep., North Macedonia, Poland, Romania, Slovakia, Slovenia, Turkey, Ukraine;
  \item North ($N_3 = 6$): Denmark, Finland, Iceland, The Netherlands, Norway, Sweden;
\end{itemize}

For each country, CPI prices are transformed into annualized month-on-month (mom) inflation, with each inflation series sequentially cleaned of seasonal effects and outliers. The processed data can be imported using: 

\begin{CodeChunk}
\begin{CodeInput}
R> data("inflation_data", package = "FARS")
\end{CodeInput}
\end{CodeChunk}

To estimate a MLDFM through \code{mldfm()}, we first need to decide how many factors to extract from each block. We extract one global factor common to all $N$ countries, and one block-specific factor common to countries in each of the three blocks. This operation is performed as follows:

\begin{CodeChunk}
\begin{CodeInput}
R> mldfm_result <- mldfm(inflation_data, 
+  blocks = 3, 
+  block_ind = c(11, 32, 38), 
+  global = 1, 
+  local = c(1, 1, 1))
\end{CodeInput}
\end{CodeChunk}

Since we do not provide any \code{method}, \code{tol}, and \code{max_iter}, the default values are enforced. Also, by default, the dataset is standardized. The \code{mldfm} object returned is stored in the \code{mldfm_result} variable. After completion, the function \code{summary()} can be used to display an overview of the estimated MLDFM, including the number of factors extracted at each level of the hierarchical structure used in the SLS estimation.

\begin{CodeChunk}
\begin{CodeInput}
R> summary(mldfm_result)
\end{CodeInput}
\begin{CodeOutput}
Summary of Multilevel Dynamic Factor Model (MLDFM)
===================================================
Number of periods               : 239 
Number of factors               : 4 
Number of nodes                 : 4 
Initialization method           : CCA 
Number of iterations to converge: 33 

Factor structure:
 - 1-2-3 :  1 factor(s)
 - 1 :  1 factor(s)
 - 2 :  1 factor(s)
 - 3 :  1 factor(s)

Residual diagnostics:
 - Total residual sum of squares (RSS):  4471.70 
 - Average RSS per time period        :  18.71 
\end{CodeOutput}
\end{CodeChunk}

\begin{figure}[!h]
  \centering
  \begin{subfigure}{0.48\linewidth}
    \centering
    \includegraphics[width=\linewidth]{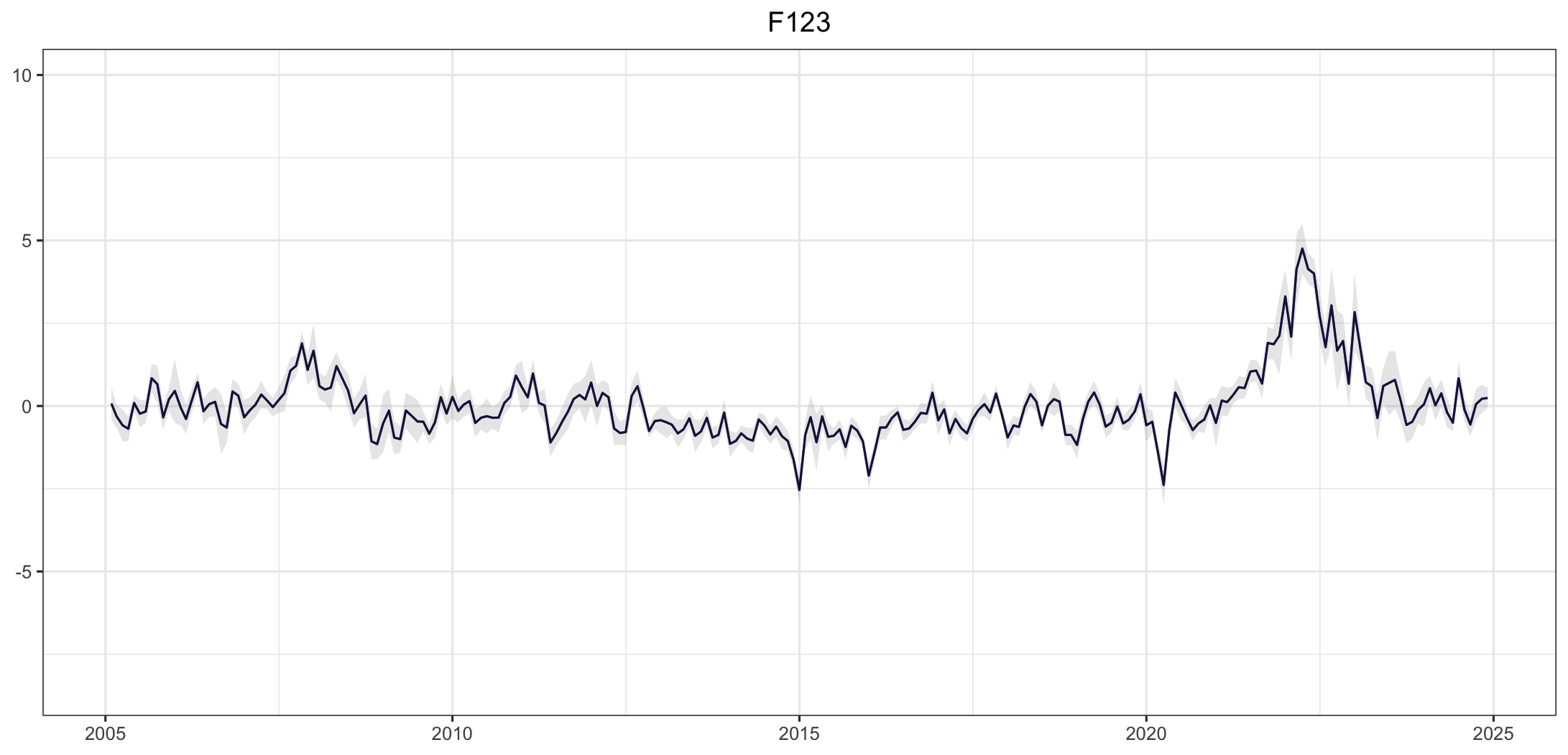}
    \caption{Global factor}
    \label{fig:f123}
  \end{subfigure}
  \hfill
  \begin{subfigure}{0.48\linewidth}
    \centering
    \includegraphics[width=\linewidth]{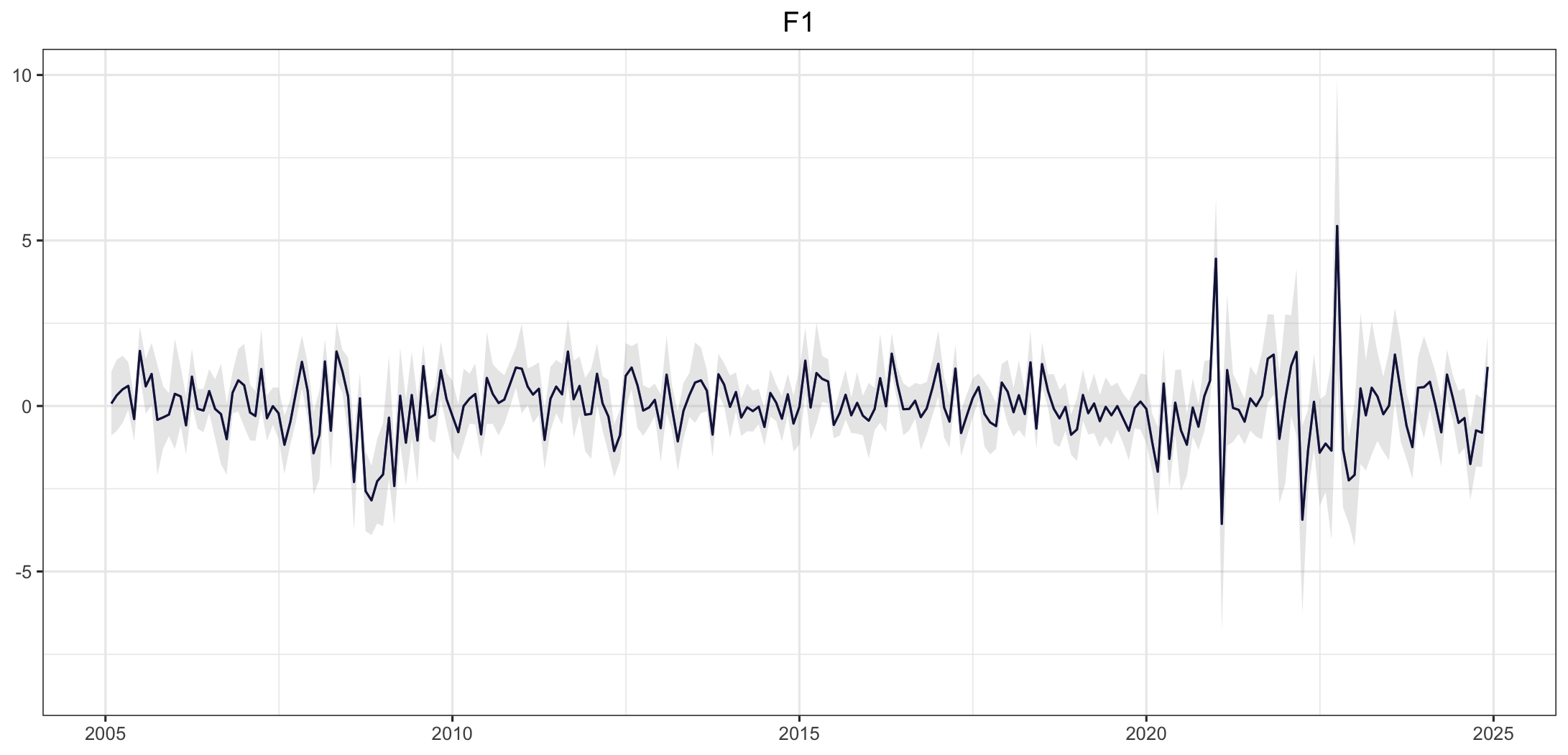}
    \caption{Block 1 factor}
    \label{fig:f1}
  \end{subfigure}
  
\vspace{1em}
  
  \begin{subfigure}{0.48\linewidth}
    \centering
    \includegraphics[width=\linewidth]{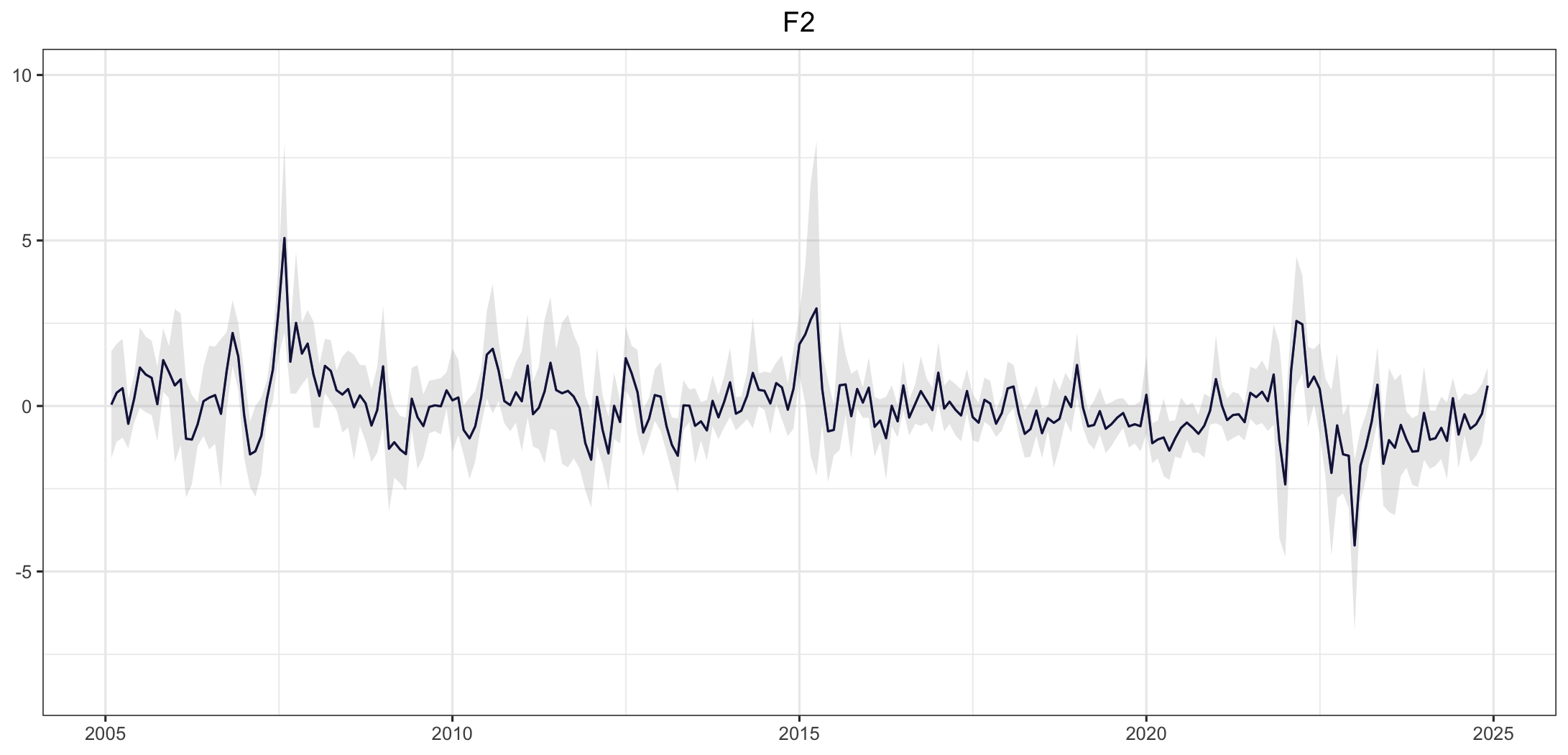}
    \caption{Block 2 factor}
    \label{fig:f2}
  \end{subfigure}
  \hfill
  \begin{subfigure}{0.48\linewidth}
    \centering
    \includegraphics[width=\linewidth]{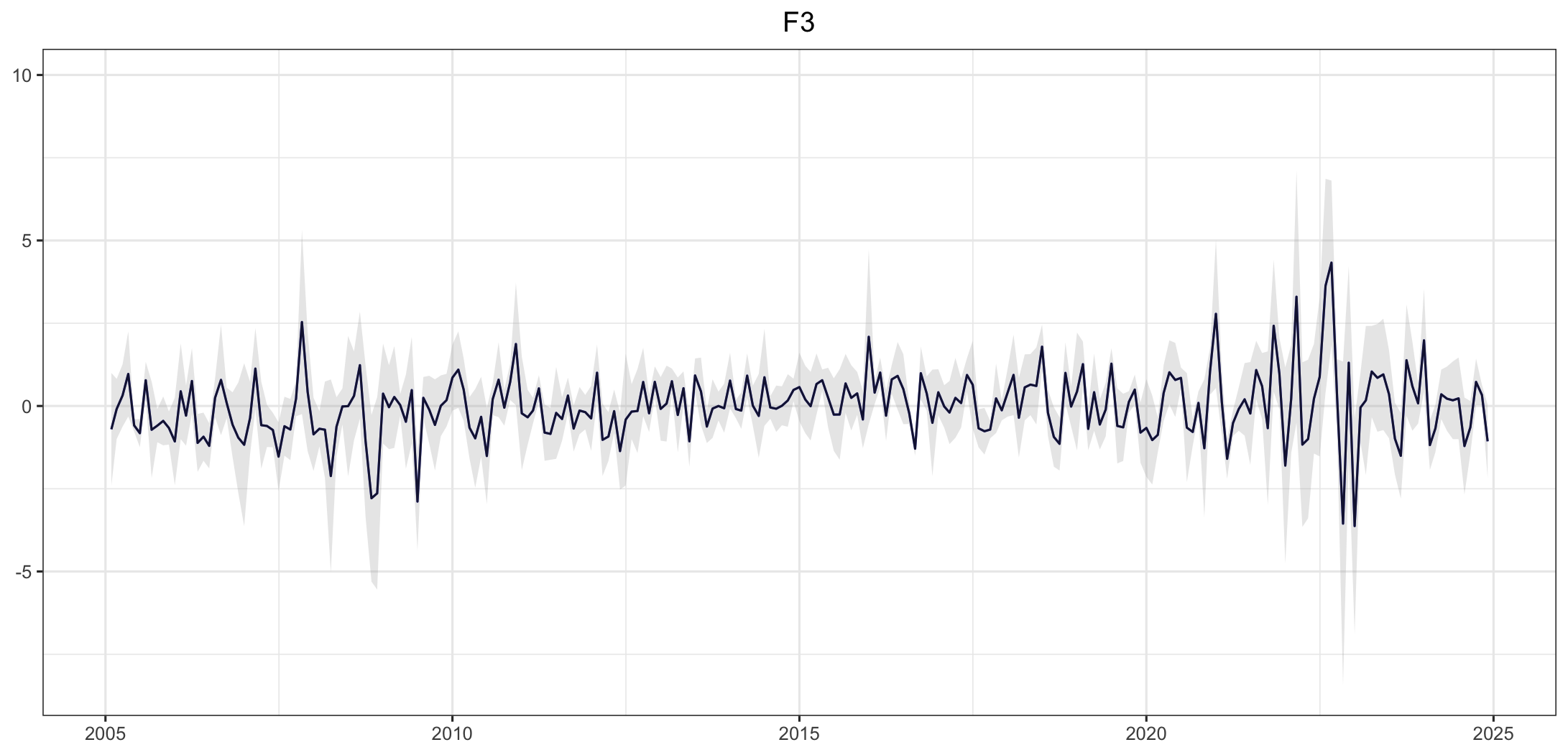}
    \caption{Block 3 factor}
    \label{fig:f3}
  \end{subfigure}
  
  \caption{Estimated factors of headline inflation in Europe together with 95\% confidence bounds.}
  \label{fig:factors}
\end{figure}

\begin{figure}[!h]
  \centering
  \begin{subfigure}{0.45\linewidth}
    \centering
    \includegraphics[width=\linewidth]{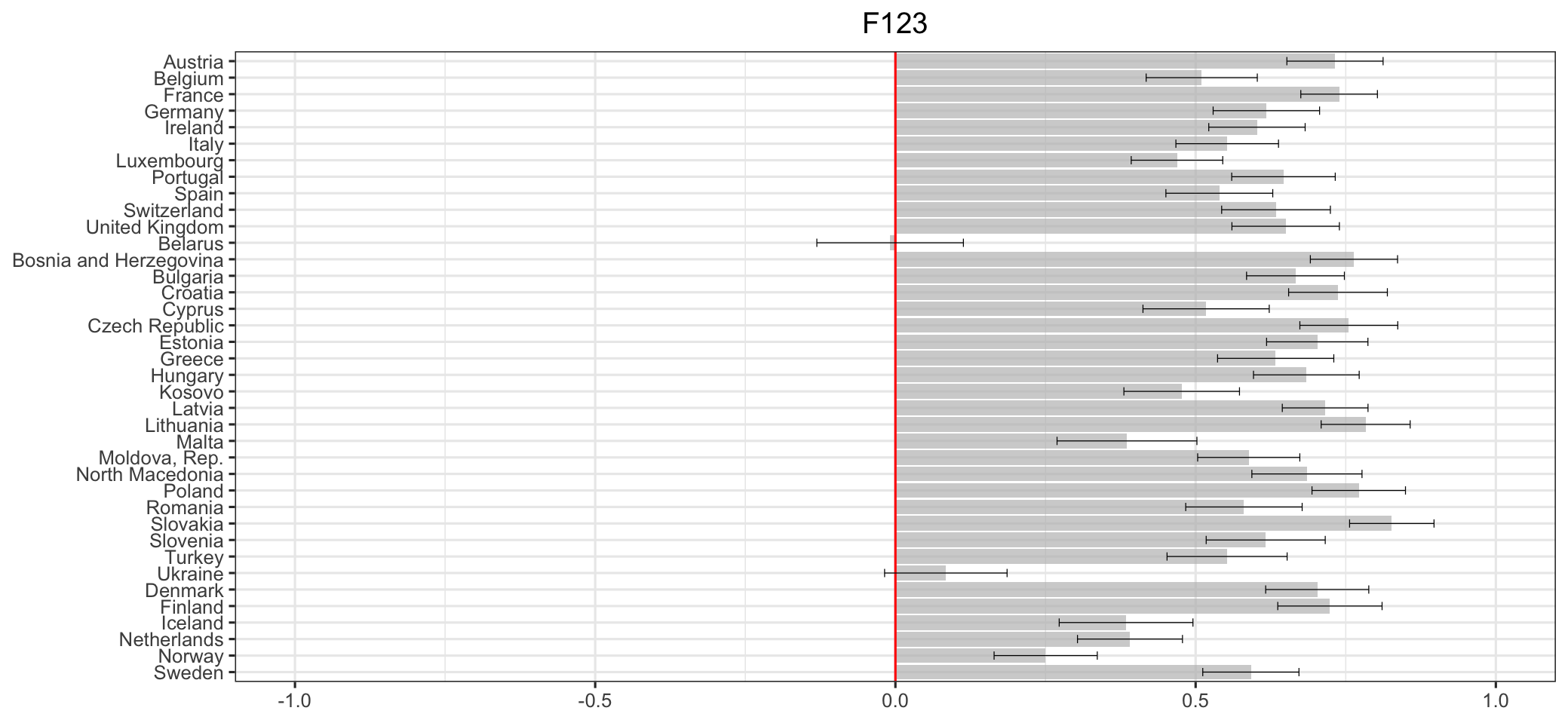}
    \caption{Global factor loadings}
    \label{fig:l123}
  \end{subfigure}
  \hfill
  \begin{subfigure}{0.45\linewidth}
    \centering
    \includegraphics[width=\linewidth]{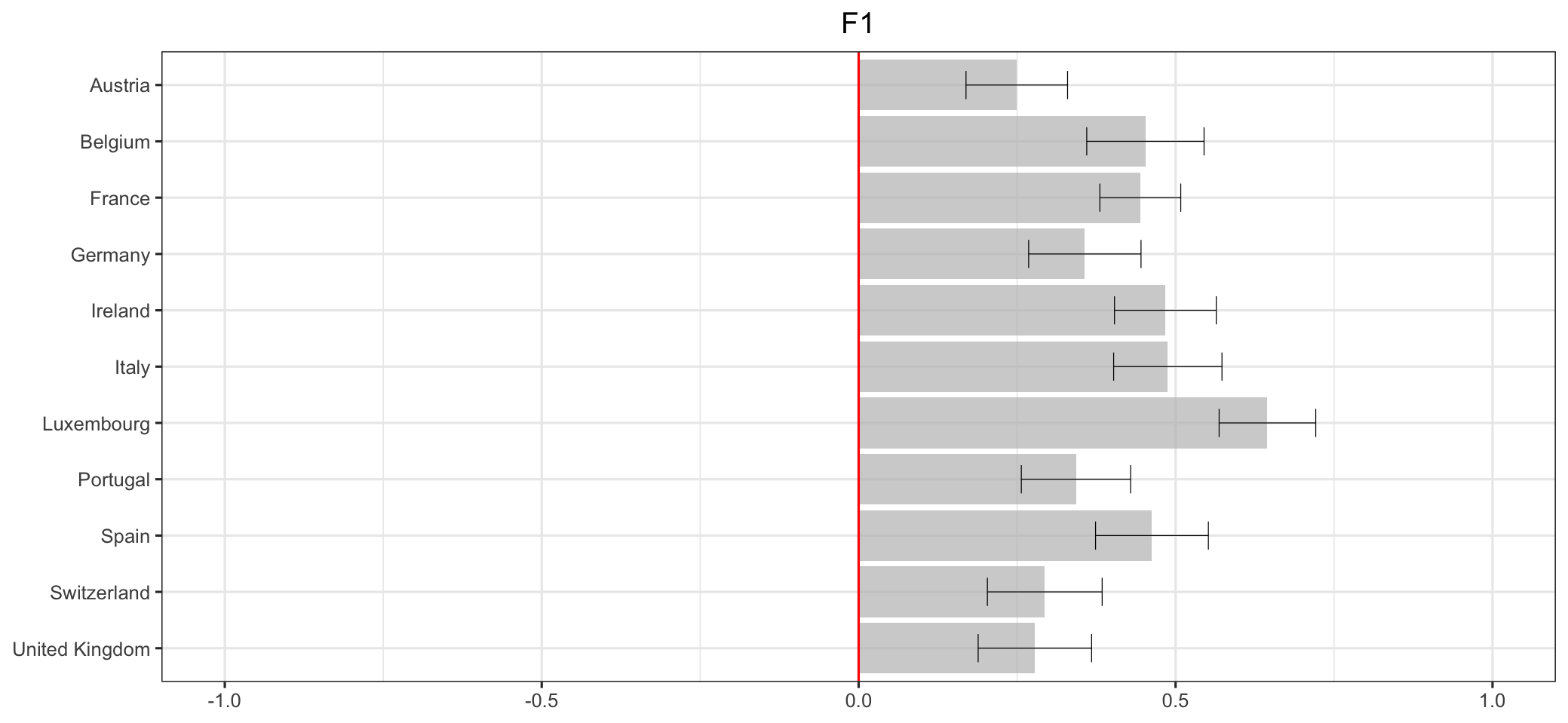}
    \caption{Block 1 factor loadings}
    \label{fig:l1}
  \end{subfigure}
  
  \vspace{1em}
  
  \begin{subfigure}{0.45\linewidth}
    \centering
    \includegraphics[width=\linewidth]{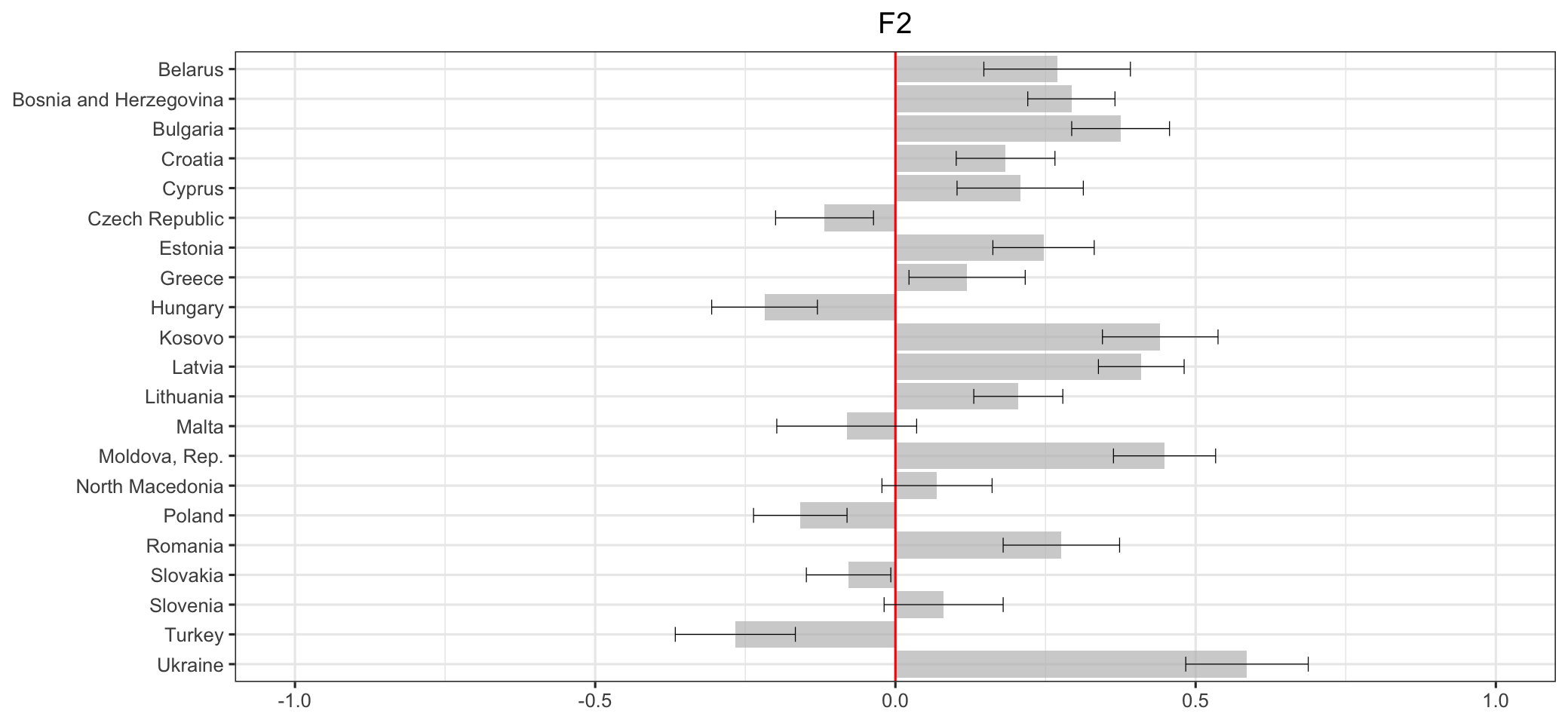}
    \caption{Block 2 factor loadings}
    \label{fig:l2}
  \end{subfigure}
  \hfill
  \begin{subfigure}{0.45\linewidth}
    \centering
    \includegraphics[width=\linewidth]{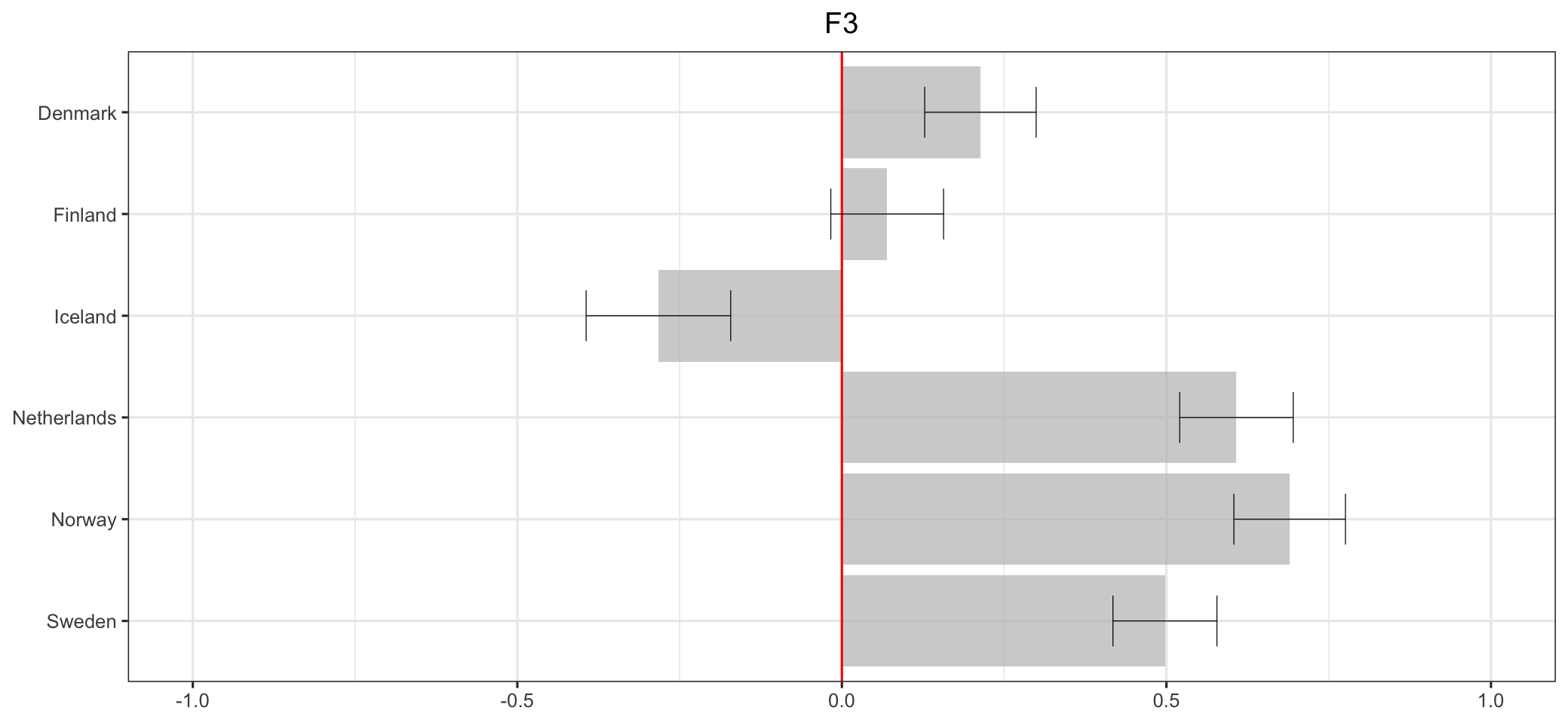}
    \caption{Block 3 factor loadings}
    \label{fig:l3}
  \end{subfigure}
  
  \caption{Estimated factor loadings of headline inflation in Europe together with 95\% confidence bounds.}
  \label{fig:loadings}
\end{figure}

\begin{figure}[!h]
    \centering
    \includegraphics[width=0.48\textwidth]{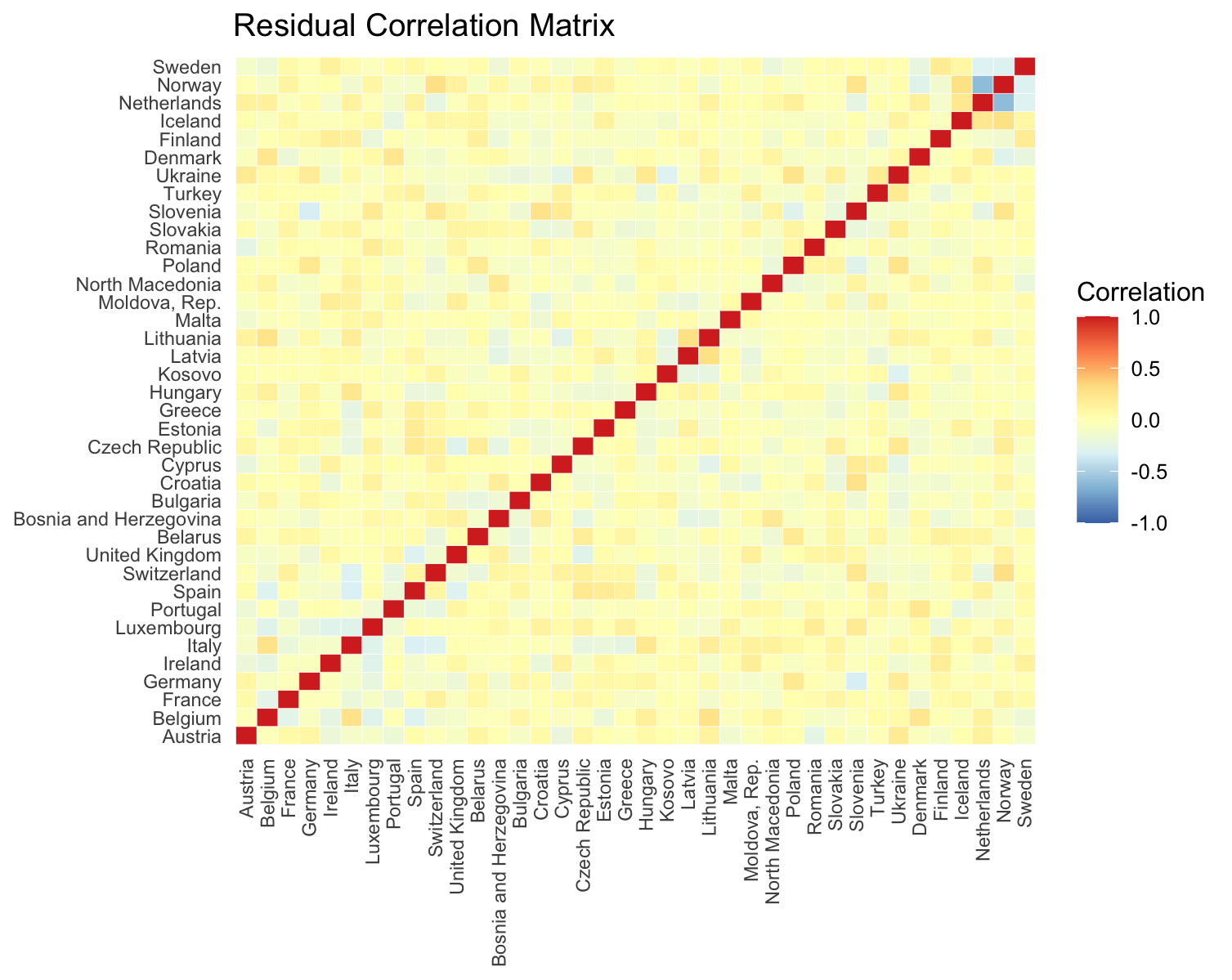}
    \caption{Correlation heatmap of estimated idiosyncratic components of headline inflation in Europe.}
    \label{fig:residuals}
\end{figure}

Additionally, using \code{plot()}, it is possible to obtain a graphical representation of the estimated factors, loadings, and residuals. This is performed by calling the plot function three times in sequence. For a more precise result, we provide the plot function with appropriate arrays composed of dates and country names using the optional arguments. Also, we specify that the global factor must be flipped in sign. The results are plotted in Figures \ref{fig:factors},~\ref{fig:loadings} and~\ref{fig:residuals}, respectively. 

\begin{CodeChunk}
\begin{CodeInput}
R> plot(mldfm_result, dates = dates, flip = c(1, 0, 0, 0))
R> plot(mldfm_result, which = "loadings", var_names = countries, 
+  flip = c(1, 0, 0, 0))
R> plot(mldfm_result, which = "residuals", var_names = countries)
\end{CodeInput}
\end{CodeChunk}

In order to analyze potential inflation risk in Europe we utilize Germany as an example. To do this, we extract the corresponding inflation series from the data set and the factors from \code{mldfm_result}.

\begin{CodeChunk}
\begin{CodeInput}
R> dep_variable <- as.numeric(inflation_data[[4]])
R> factors <- factors(mldfm_result)
\end{CodeInput}
\end{CodeChunk}

The next step is to estimate the FA-QRs which will be used as baseline for our scenario analysis:\footnote{For this task, we consider the simplest case with \code{h = 1}.}

\begin{CodeChunk}
\begin{CodeInput}
R> fars_result <- compute_fars(dep_variable, 
+  factors, 
+  h = 1) 
\end{CodeInput}
\begin{CodeOutput}
Running Factor-Augmented Quantile Regressions (FA-QRs)...
Completed
\end{CodeOutput}
\end{CodeChunk}

After this, we can print a recap of the FA-QRs and inspect the estimated coefficients across all quantile regressions. Additional information can be observed calling \code{summary()} function. 
\begin{CodeChunk}
\begin{CodeInput}
R> print(fars_result)
\end{CodeInput}
\begin{CodeOutput}
Factor-Augmented Quantile Regressions (FARS)
===========================================
Number of periods:  238 
Number of factors:  4 
Lag:  1 
Quantile levels:  0.05 0.25 0.50 0.75 0.95 
\end{CodeOutput}
\begin{CodeInput}
R> coef(fars_result)
\end{CodeInput}
\begin{CodeOutput}
              0.05   0.25   0.50   0.75   0.95
(Intercept) -2.554  0.030  2.066  3.475  6.964
LagY         0.096  0.046  0.024 -0.013 -0.215
F1          -0.909 -0.889 -1.216 -1.420 -2.531
F2           0.962  0.220 -0.216 -0.139 -0.211
F3           1.029  0.435  0.217  0.102  0.164
F4          -0.093  0.340  0.370  0.256  1.086
\end{CodeOutput}
\end{CodeChunk}

If the user is interested in performing additional diagnostic analyses on a specific quantile, the corresponding \code{rq} object can be retrieved as follows:

\begin{CodeChunk}
\begin{CodeInput}
R> get_rq_model(fars_result, tau = 0.25)
\end{CodeInput}
\begin{CodeOutput}
Call:
quantreg::rq(formula = y ~ ., tau = tau_i, data = df)

Coefficients:
(Intercept)        LagY          F1          F2          F3          F4 
 0.03043957  0.04618208 -0.88931471  0.22043590  0.43533163  0.34005343 

Degrees of freedom: 238 total; 232 residual
\end{CodeOutput}
\end{CodeChunk}

\subsubsection{Non-stressed scenario}
\begin{figure}[!h]
  \centering
  \begin{subfigure}{0.48\linewidth}
    \centering
    \includegraphics[width=\linewidth]{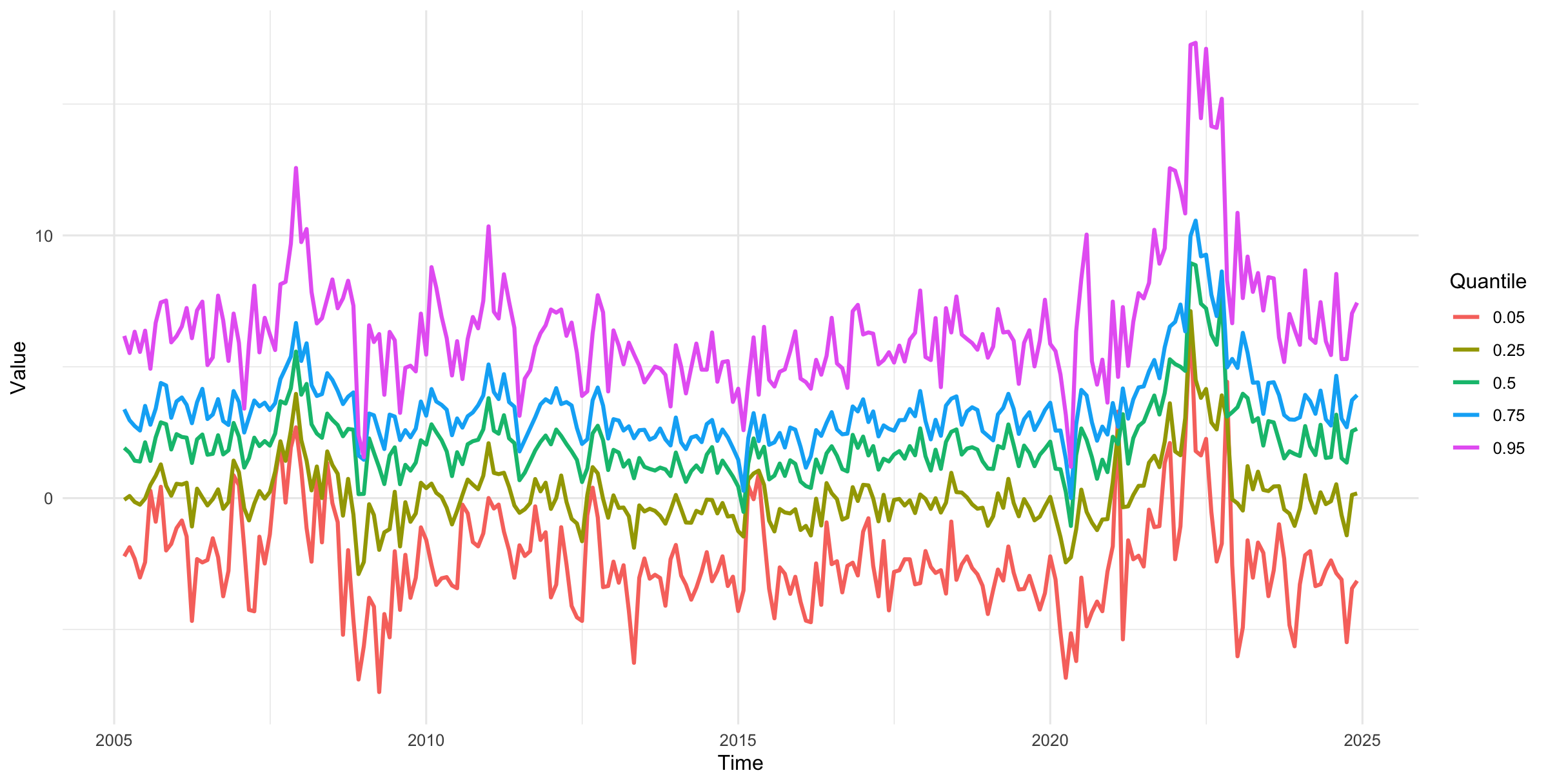}
    \caption{Non-stressed scenario}
    \label{fig:Inf_Q}
  \end{subfigure}
  \hfill 
  \begin{subfigure}{0.48\linewidth}
    \centering
    \includegraphics[width=\linewidth]{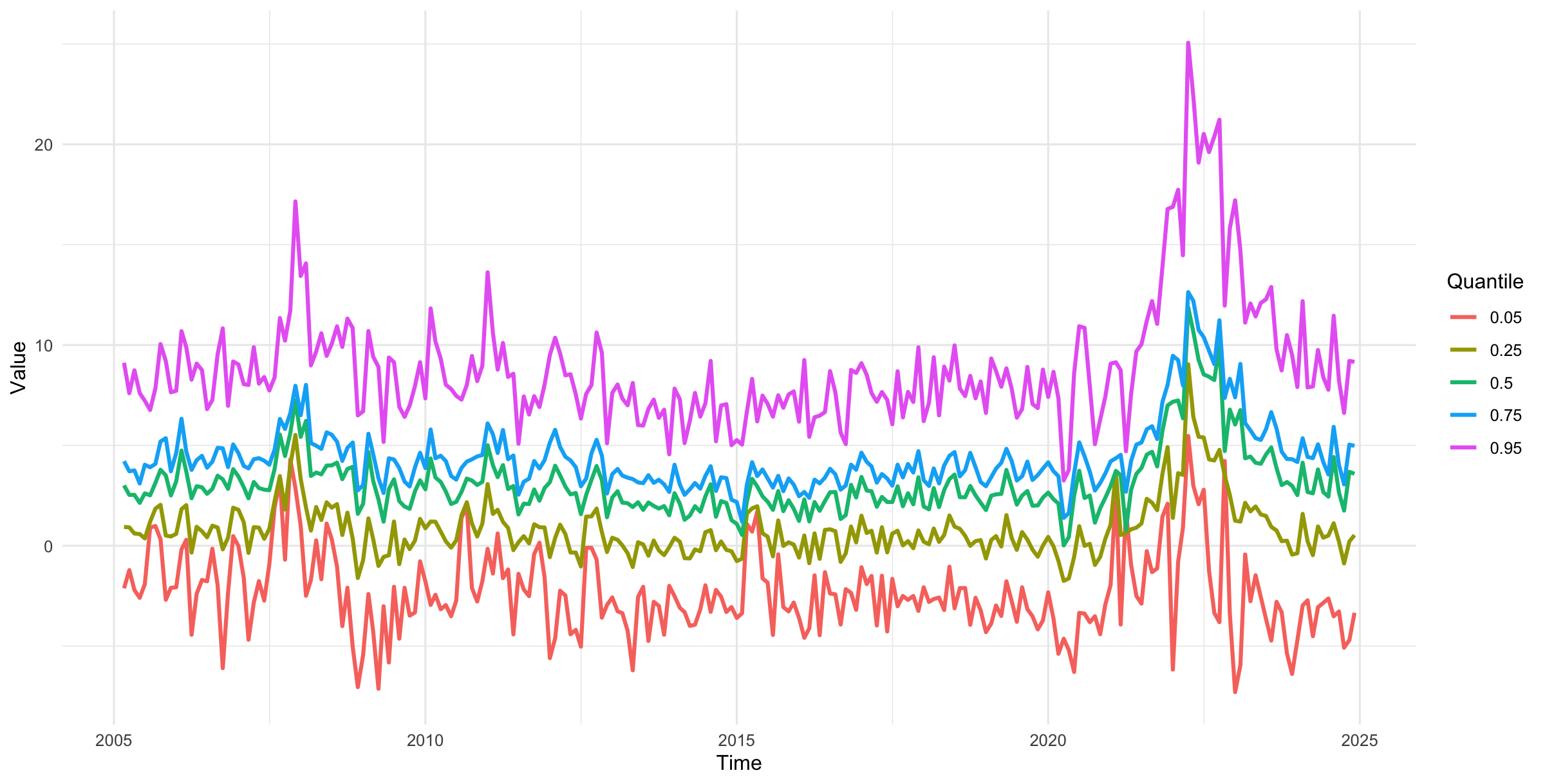}
    \caption{Stressed scenario}
    \label{fig:Inf_S_Q}
  \end{subfigure}
  \caption{Non-stressed (left panel) and stressed scenario (right panel) quantiles for Germany headline inflation.}
  \label{fig:inflation_quantiles}
\end{figure}

To estimate the density under the non-stressed scenario we first need the quantiles' series. For this example, we use the in-sample estimates obtained with the \code{fitted()} method of \code{fars} object.\footnote{The same result could be obtained using \code{predict()} and passing as \code{newdata} the same data used for fitting the FA-QRs.} The fitted values can be viasualized using the plot function (see Figure~\ref{fig:inflation_quantiles}, panel a). 

\begin{CodeChunk}
\begin{CodeInput}
R> quantiles <- fitted(fars_result)
R> plot(fars_result, dates = dates[-1])
\end{CodeInput}
\end{CodeChunk}

The estimates, stored in the \code{quantiles} variable, are then used to fit a skew-t distribution, generating the density for the non-stressed scenario. This is done by applying the non-linear optimization method and providing an appropriate support for the inflation case.

\begin{CodeChunk}
\begin{CodeInput}
R> ns_density <- compute_density(quantiles, 
+  support = c(-30, 30), 
+  seed = 42, 
+  nl = TRUE)
\end{CodeInput}
\begin{CodeOutput}
Estimating skew-t densities from forecasted quantiles...
Completed
\end{CodeOutput}
\end{CodeChunk}

\begin{figure}[!h]
  \centering
  \begin{subfigure}{0.48\linewidth}
    \centering
    \includegraphics[width=\linewidth]{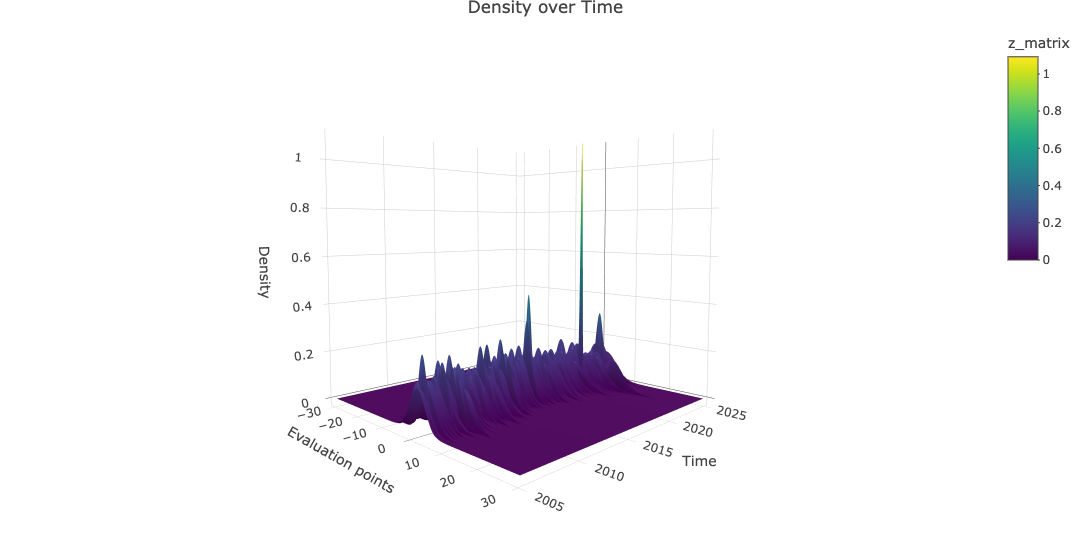}
    \caption{Non-stressed density}
    \label{fig:Inf_D}
  \end{subfigure}
  \hfill
  \begin{subfigure}{0.48\linewidth}
    \centering
    \includegraphics[width=\linewidth]{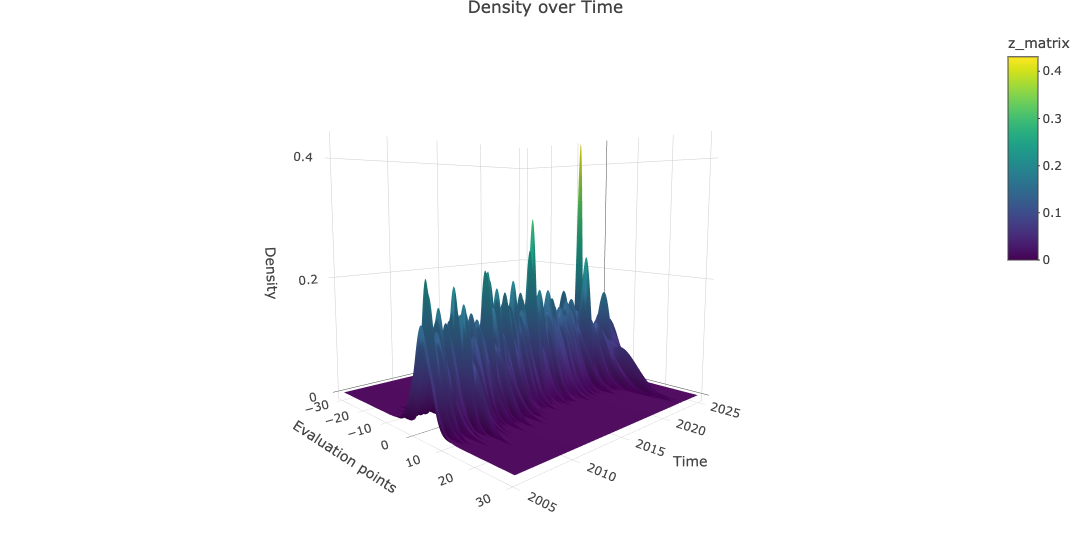}
    \caption{Stressed density}
    \label{fig:Inf_S_D}
  \end{subfigure}
  \caption{Non-stressed (left panel) and stressed (right panel) densities for Germany headline inflation.}
  \label{fig:inflation_densities}
\end{figure}

The generated \code{fars_density} object can be used to plot the non-stressed density (see Figure~\ref{fig:inflation_densities}, panel a) and visualize an overview of the density estimation.

\begin{CodeChunk}
\begin{CodeInput}
R> plot(ns_density, time_index = dates[-1])
R> print(ns_density)
\end{CodeInput}
\begin{CodeOutput}
FARS Density
====================
Time observations  : 238 
Estimation points  : 512 
Random samples     : 5000 
Support range      : [ -30 , 30 ]
Optimization       : Non-linear 
\end{CodeOutput}
\end{CodeChunk}

Finally, we derive the inflation at risk (IaR) at qtau = 0.99 applying the \code{quantile_risk()} function to the non-stressed density.

\begin{CodeChunk}
\begin{CodeInput}
R> IaR <- quantile_risk(ns_density, qtau = 0.99)
\end{CodeInput}
\end{CodeChunk}

\subsubsection{Stressed scenarios}

As explained in Section~\ref{sec:FARS_package}, the computation of stressed scenarios can be performed in two steps. First, we need to obtain the asymptotic distribution of the factors. For this goal, we implement the subsampling procedure using the appropriate function. In our case, we generate 100 samples by extracting $95\%$ of the countries in each block. 

\begin{CodeChunk}
\begin{CodeInput}
R> mldfm_ss_result <- mldfm_subsampling(inflation_data,
+  blocks = 3, 
+  block_ind = c(11, 32, 38), 
+  global = 1,
+  local = c(1, 1, 1), 
+  n_samples = 100, 
+  sample_size = 0.95,
+  seed = 42)
\end{CodeInput}
\begin{CodeOutput}
Generating 100 subsamples...
Subsampling completed.
\end{CodeOutput}
\end{CodeChunk}

Each of the 100 models stored in \code{mldfm_ss_result} can be manipulated as a distinct \code{mldfm} object. For example, we can visualize the summary of the MLDFM estimated for sample number 10.

\begin{CodeChunk}
\begin{CodeInput}
R> summary(get_mldfm_model(mldfm_ss_result, index = 10))
\end{CodeInput}
\begin{CodeOutput}
Summary of Multilevel Dynamic Factor Model (MLDFM)
===================================================
Number of periods               : 239 
Number of factors               : 4 
Number of nodes                 : 4 
Initialization method           : CCA 
Number of iterations to converge: 23 

Factor structure:
 - 1-2-3 :  1 factor(s)
 - 1 :  1 factor(s)
 - 2 :  1 factor(s)
 - 3 :  1 factor(s)

Residual diagnostics:
 - Total residual sum of squares (RSS):  4197.49 
 - Average RSS per time period        :  17.56 
\end{CodeOutput}
\end{CodeChunk}

The second step is to generate the stressed scenario by calling the \code{create_scenario()} function. For this exercise, we consider the highest stress level of \code{alpha = 0.99} and default $\hat{\Gamma}_t^{\text{BN}}$.

\begin{CodeChunk}
\begin{CodeInput}
R> scenario <- create_scenario(model = mldfm_result, 
+  subsample = mldfm_ss_result, 
+  alpha = 0.99)
\end{CodeInput}
\begin{CodeOutput}
Constructing scenario using 100 subsamples, alpha = 0.99 
and standard time-varying Gamma...
Scenario construction completed.
\end{CodeOutput}
\end{CodeChunk}

A summary of the scenario can be displayed as follows:
\begin{CodeChunk}
\begin{CodeInput}
R> summary(scenario)
\end{CodeInput}
\begin{CodeOutput}
FARS Scenario Summary
======================
Number of periods    : 239 
Ellipsoid dimensions : 4 
Points per ellipsoid : 1072 
Confidence level     : 99 %
FPR Gamma            : FALSE 

Center (factor estimates):
  Mean     : 0 
  Std. Dev : 1.0005 
  Min      : -4.7547 
  Max      : 5.4379 

Ellipsoid variability (diagonal of Sigma):
  Mean     : 0.3285 
  Std. Dev : 0.5681 
  Min      : 0.0079 
  Max      : 7.2701 
\end{CodeOutput}
\end{CodeChunk}

Now that we have our scenario we can stress accordingly our factors. Since we are interested in inflation risk, our objective is to maximize the dependent variable for the chosen quantile (\code{qtau = 0.99}). 

\begin{CodeChunk}
\begin{CodeInput}
R> ellipsoids <- get_ellipsoids(scenario)
R> stressed_factors <- compute_stressed_factors(dep_variable, 
+  factors, 
+  h = 1, 
+  qtau = 0.99, 
+  direction = "max", 
+  ellipsoids = ellipsoids)
\end{CodeInput}
\begin{CodeOutput}
\end{CodeOutput}
\end{CodeChunk}

Using the new factors, the stressed quantiles can be estimated with the \code{predict()} method of our \code{fars} object.\footnote{The last row of the \code{newdata} matrix is dropped to keep only the in-sample period.} Moreover, by providing \code{newdata} to the \code{plot()} function, it is also possible to visualize the stressed quantiles (see Figure~\ref{fig:inflation_quantiles} panel b).

\begin{CodeChunk}
\begin{CodeInput}
R> newdata <- cbind(dep_variable, stressed_factors)
R> newdata <- newdata[-nrow(newdata), ]
R> stressed_quantiles <- predict(fars_result, newdata)
R> plot(fars_result, newdata, dates[-1])
\end{CodeInput}
\begin{CodeOutput}
\end{CodeOutput}
\end{CodeChunk}

As in the non-stressed case, we fit a skew-t distribution to obtain the stressed density, this time using the matrix of stressed quantiles as input. Again, we can visualize the density with the \code{plot} function (see Figure~\ref{fig:inflation_densities}, panel b).

\begin{CodeChunk}
\begin{CodeInput}
R> s_density <- compute_density(stressed_quantiles, 
+  support = c(-30, 30), 
+  seed = 42, 
+  nl = TRUE)
\end{CodeInput}
\begin{CodeOutput}
Estimating skew-t densities from forecasted quantiles...
Completed
\end{CodeOutput}
\end{CodeChunk}

The last step is deriving the inflation in stress (IiS) for \code{qtau = 0.99} by feeding \code{quantile_risk()} with the stressed densities.

\begin{CodeChunk}
\begin{CodeInput}
R> IiS <- quantile_risk(s_density, qtau = 0.99)
\end{CodeInput}
\end{CodeChunk}

\begin{figure}[!h]
    \centering
    \includegraphics[width=0.7\textwidth]{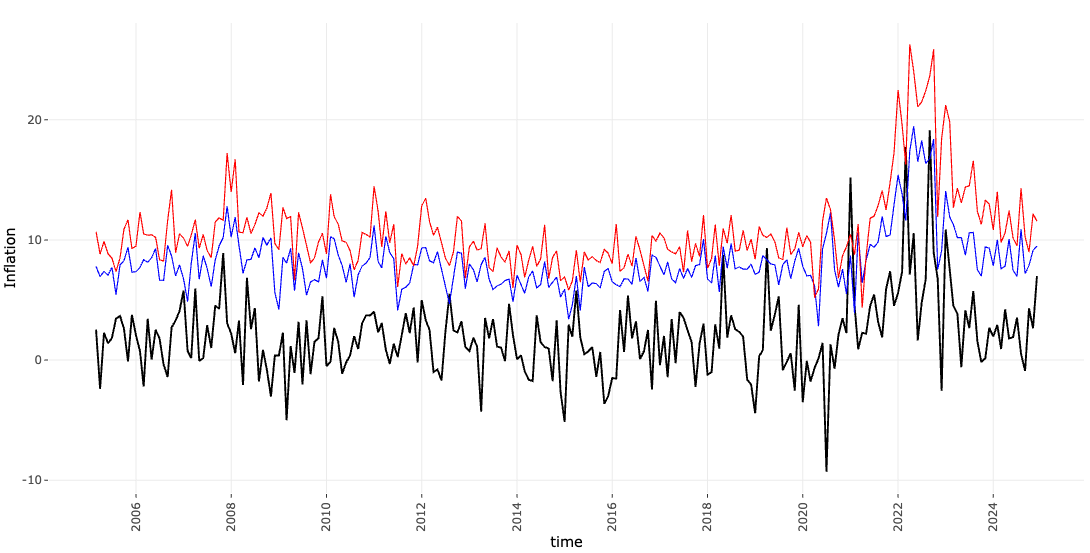}
    \caption{Germany monthly mom headline inflation (black), together with 99\% IaR (blue) and 99\% IiS stressed with $\alpha$ = 99\% (red).}
    \label{fig:inflation_final_plot}
\end{figure}

In Figure~\ref{fig:inflation_final_plot}, we plot the final IaR and IiS estimates along with the dependent variables for the period spanning from March 2005 to December 2024. We observe that IiS is higher than IaR. This worse outcome would be neglected if we only estimated IaR, which assumes that factors evolve according to an average scenario.

\subsection{Measuring growth risk in the United States using FARS}
\label{sec:real_example}

In our second illustration, we follow \cite{Gonzlez-Rivera2024} and construct densities for annualized quarterly GDP growth in the US with the underlying factors extracted in the context of a MLDFM using a data sample composed of three blocks. The first block contains $N_1 = 63$ international macroeconomic variables (GDP growth for 63 countries), the second block contains $N_2 = 248$ domestic macroeconomic variables, and the third block contains $N_3 = 208$ international financial variables. All variables are observed quarterly from 2005Q3 to 2020Q1.\footnote{Data are retrieved from the replication files of \cite{Gonzlez-Rivera2024}.} The dataset, composed of $N = N_1 + N_2 + N_3 = 519$ variables and the US GDP growth can be imported using: 

\begin{CodeChunk}
\begin{CodeInput}
R> data("mf_data", package = "FARS")
R> data("dep_variable", package = "FARS")
\end{CodeInput}
\end{CodeChunk}

We extract one global factor common to all $N$ variables, a pairwise factor common to all international variables (international macroeconomic and international financial blocks), and one block-specific factor common to the variables in each of the three blocks. Then, we check the summary of the model.

\begin{CodeChunk}
\begin{CodeInput}
R> mldfm_result <- mldfm(mf_data, 
+  blocks = 3, 
+  block_ind = c(63, 311, 519), 
+  global = 1, 
+  local = c(1, 1, 1), 
+  middle_layer = list("1-3" = 1))
R> summary(mldfm_result)
\end{CodeInput}
\begin{CodeOutput}
Summary of Multilevel Dynamic Factor Model (MLDFM)
===================================================
Number of periods               : 59 
Number of factors               : 5 
Number of nodes                 : 5 
Initialization method           : CCA 
Number of iterations to converge: 47 

Factor structure:
 - 1-2-3 :  1 factor(s)
 - 1-3 :  1 factor(s)
 - 1 :  1 factor(s)
 - 2 :  1 factor(s)
 - 3 :  1 factor(s)

Residual diagnostics:
 - Total residual sum of squares (RSS):  14770.03 
 - Average RSS per time period        :  250.34 
\end{CodeOutput}
\end{CodeChunk}

Using the extracted factors, we fit our FA-QR models, this time changing the extreme quantiles to 1\% and 99\%.

\begin{CodeChunk}
\begin{CodeInput}
R> factors <- factors(mldfm_result)
R> fars_result <- compute_fars(dep_variable, 
+  factors, 
+  h = 1, 
+  edge = 0.01) 
\end{CodeInput}
\begin{CodeOutput}
Running Factor-Augmented Quantile Regressions (FA-QRs)...
Completed
\end{CodeOutput}
\begin{CodeInput}
R> print(fars_result)
\end{CodeInput}
\begin{CodeOutput}
Factor-Augmented Quantile Regressions (FARS)
===========================================
Number of periods:  58 
Number of factors:  5 
Lag:  1 
Quantile levels:  0.01 0.25 0.50 0.75 0.99 
\end{CodeOutput}
\end{CodeChunk}

\subsubsection{Non-stressed scenario}
\begin{figure}[!h]
  \centering
  \begin{subfigure}{0.48\linewidth}
    \centering
    \includegraphics[width=\linewidth]{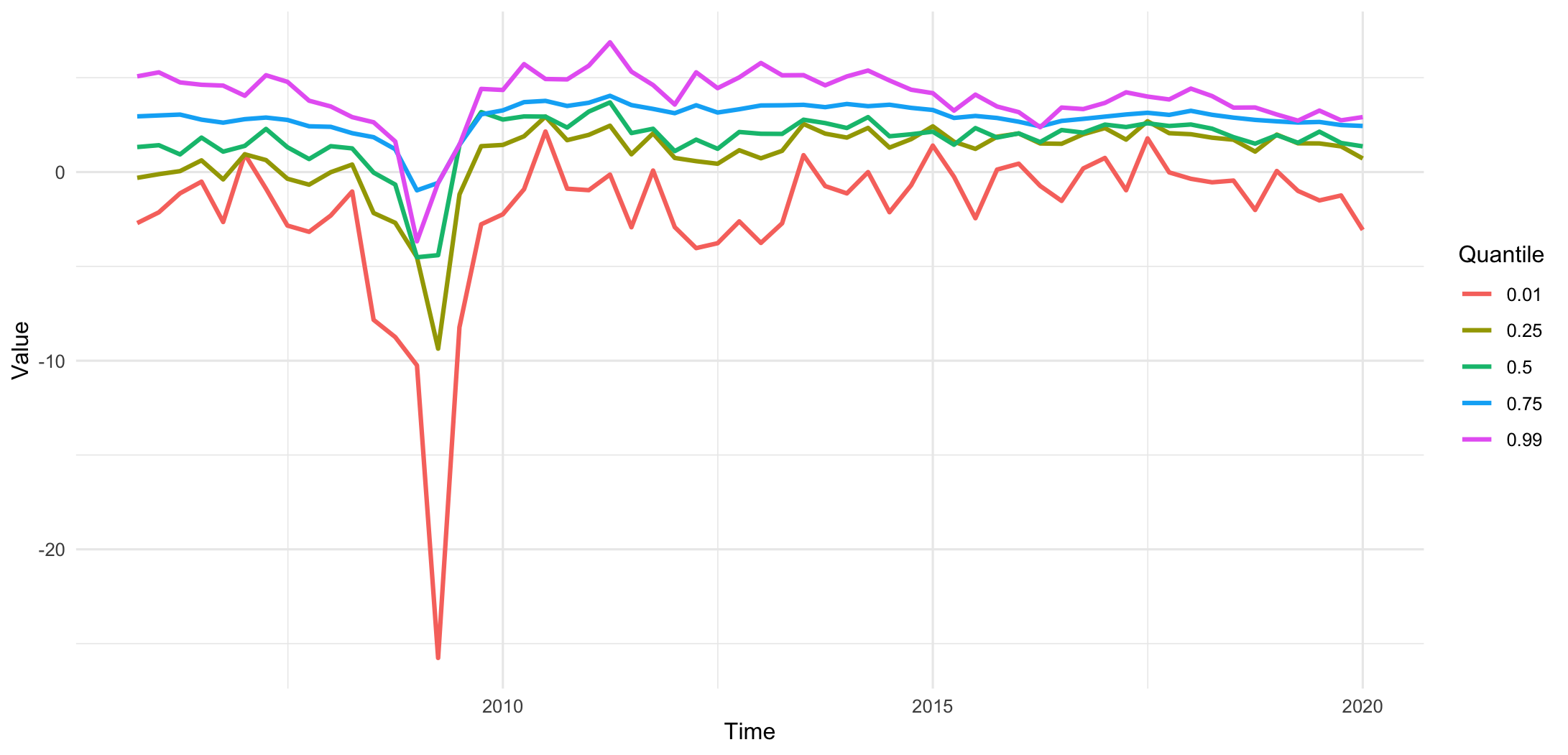}
    \caption{Non-stressed scenario}
    \label{fig:GDP_Q}
  \end{subfigure}
  \hfill 
  \begin{subfigure}{0.48\linewidth}
    \centering
    \includegraphics[width=\linewidth]{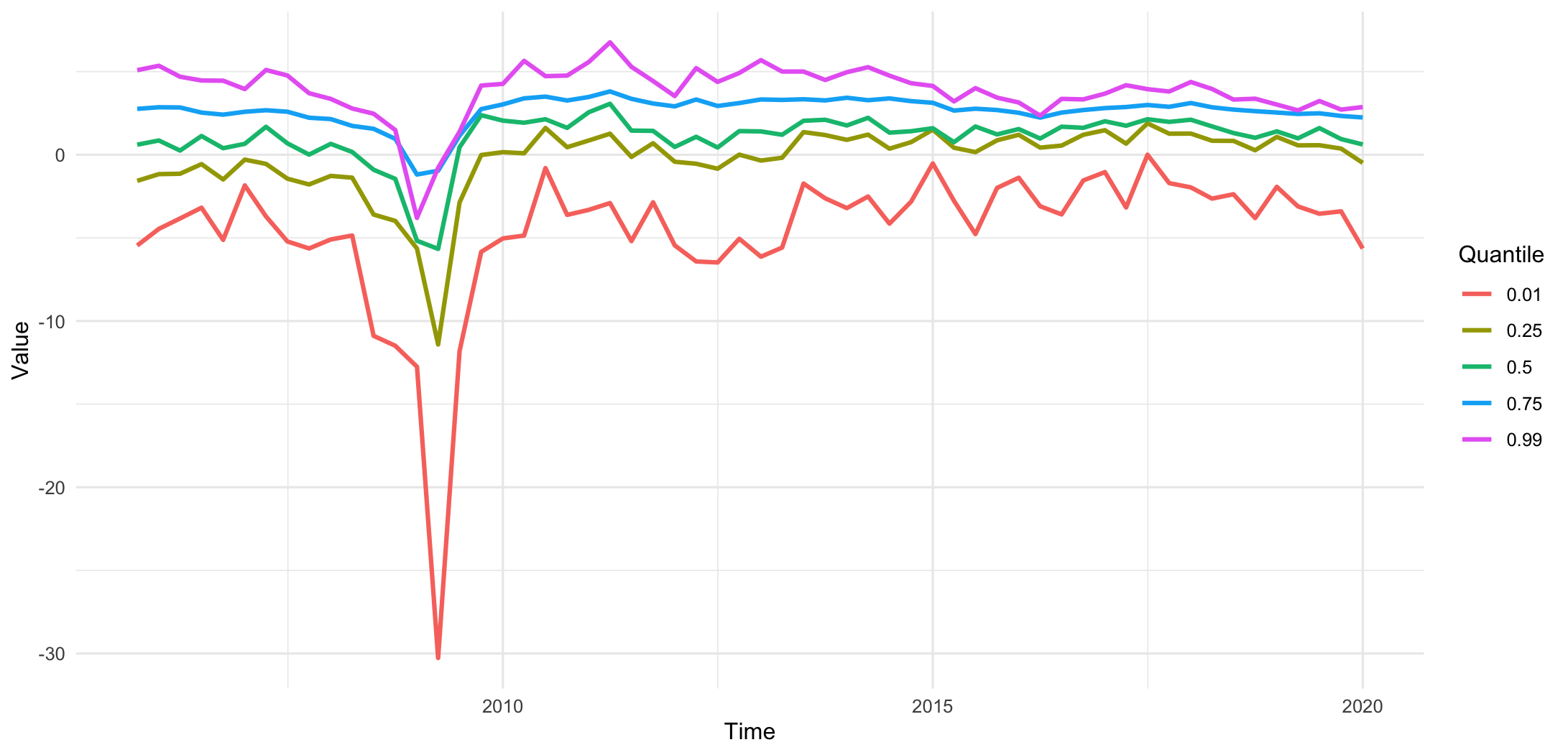}
    \caption{Stressed scenario}
    \label{fig:GDP_S_Q}
  \end{subfigure}
  \caption{Non-stressed (left panel) and stressed scenario (right panel) quantiles for US GDP growth.}
  \label{fig:gdp_quantiles}
\end{figure}

\begin{figure}[!h]
  \centering
  \begin{subfigure}{0.48\linewidth}
    \centering
    \includegraphics[width=\linewidth]{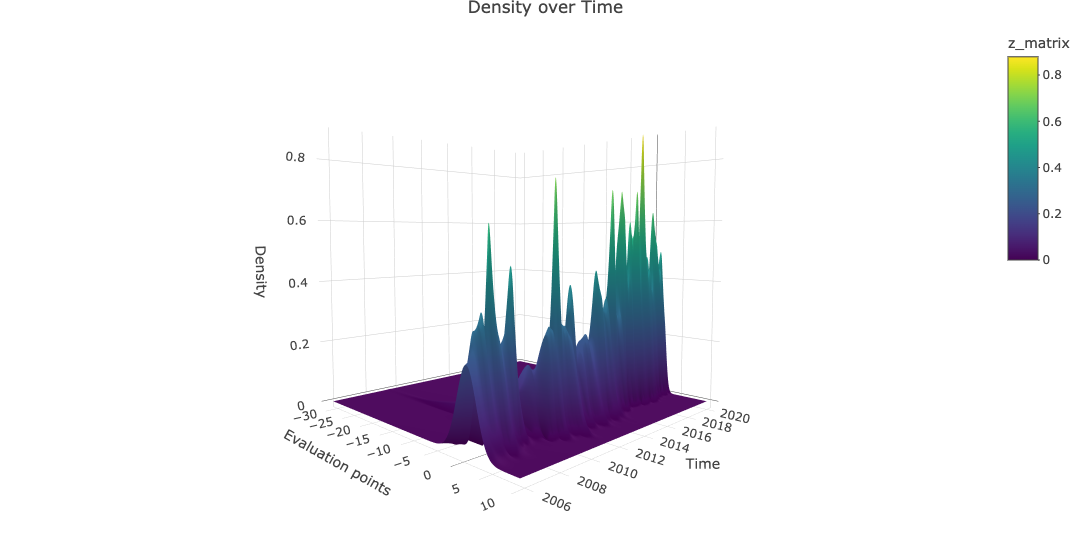}
    \caption{Non-stressed density}
    \label{fig:GDP_D}
  \end{subfigure}
  \hfill
  \begin{subfigure}{0.48\linewidth}
    \centering
    \includegraphics[width=\linewidth]{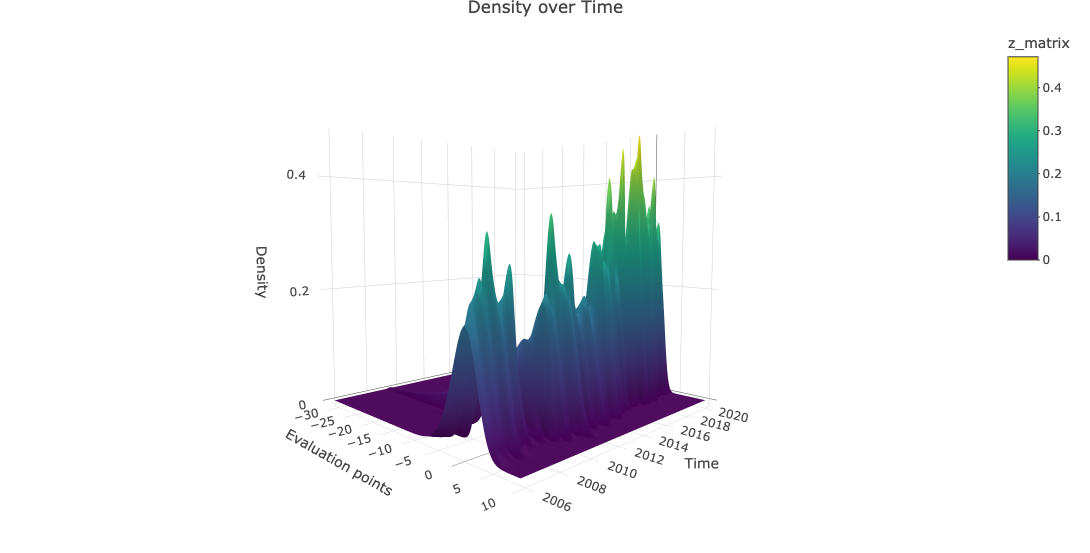}
    \caption{Stressed density}
    \label{fig:GDP_S_D}
  \end{subfigure}
  \caption{Non-stressed (left panel) and stressed (right panel) densities for US GDP growth.}
  \label{fig:gdp_densities}
\end{figure}

As in the inflation exercise, we use the estimated quantiles to fit the non-stressed density and compute the growth at risk for the lower tail at the 1\% quantile. The quantiles and the resulting density distribution are shown in Figure~\ref{fig:gdp_quantiles}, panel a, and Figure~\ref{fig:gdp_densities}, panel a, respectively.

\begin{CodeChunk}
\begin{CodeInput}
R> quantiles <- fitted(fars_result)
R> plot(fars_result, dates = dates[-1])
\end{CodeInput}
\end{CodeChunk}

\begin{CodeChunk}
\begin{CodeInput}
R> levels <- get_quantile_levels(fars_result)
R> ns_density <- compute_density(quantiles, 
+  support = c(-30, 10), 
+  levels = levels, 
+  seed = 42)
\end{CodeInput}
\begin{CodeOutput}
Estimating skew-t densities from forecasted quantiles...
Completed
\end{CodeOutput}
\begin{CodeInput}
R> plot(ns_density, time_index = dates[-1])
R> GaR <- quantile_risk(ns_density, qtau = 0.01)
\end{CodeInput}
\begin{CodeOutput}
\end{CodeOutput}
\end{CodeChunk}

\subsubsection{Stressed scenario}
To build the stressed scenario we implement the same two-step procedure using the \code{mldfm\allowbreak_subsampling()} and \code{create_scenario()} functions. As for the inflation case, we generate 100 samples by extracting $95\%$ of the variables from each block and consider the highest stress level of \code{alpha = 0.99} with default $\hat{\boldsymbol{\Gamma}}_t^{\text{BN}}$.

\begin{CodeChunk}
\begin{CodeInput}
R> mldfm_ss_result <- mldfm_subsampling(mf_data,
+  blocks = 3, 
+  block_ind = c(63, 311, 519), 
+  global = 1,
+  local = c(1, 1, 1), 
+  middle_layer = list("1-3" = 1),
+  n_samples = 100, 
+  sample_size = 0.95, 
+  seed = 42)
\end{CodeInput}
\begin{CodeOutput}
Generating 100 subsamples...
Subsampling completed.
\end{CodeOutput}
\begin{CodeInput}
R> scenario <- create_scenario(model = mldfm_result, 
+  subsample = mldfm_ss_result, 
+  alpha = 0.99)
\end{CodeInput}
\begin{CodeOutput}
Constructing scenario using 100 subsamples, alpha = 0.99 
and standard time-varying Gamma...
Scenario construction completed.
\end{CodeOutput}
\end{CodeChunk}

Now we can stress our factors. Since we are interested in GDP growth risk, our objective is to minimize the dependent variable for the chosen low quantile (\code{qtau = 0.01}).  

\begin{CodeChunk}
\begin{CodeInput}
R> ellipsoids <- get_ellipsoids(scenario)
R> stressed_factors <- compute_stressed_factors(dep_variable, 
+  factors, 
+  h = 1, 
+  qtau = 0.01, 
+  direction = "min", 
+  ellipsoids = ellipsoids)
\end{CodeInput}
\begin{CodeOutput}
\end{CodeOutput}
\end{CodeChunk}

The stressed factors are input into the FA-QRs to estimate GDP growth quantiles under stress, from which we derive the density distribution and the corresponding growth in stress. The resulting quantiles and density distribution are shown in Figure~\ref{fig:gdp_quantiles}, panel b), and Figure~\ref{fig:gdp_densities}, panel b), respectively.

\begin{CodeChunk}
\begin{CodeInput}
R> newdata <- cbind(dep_variable, stressed_factors)
R> newdata <- newdata[-nrow(newdata), ]
R> stressed_quantiles <- predict(fars_result, newdata)
R> plot(fars_result, newdata, dates[-1])
R> s_density <- compute_density(stressed_quantiles,
+  support = c(-30, 10), 
+  levels = levels, 
+  seed = 42)
\end{CodeInput}
\begin{CodeOutput}
Estimating skew-t densities from forecasted quantiles...
Completed
\end{CodeOutput}
\begin{CodeInput}
R> plot(s_density, time_index = dates[-1])
R> GiS <- quantile_risk(s_density, qtau = 0.01)
\end{CodeInput}
\begin{CodeOutput}
\end{CodeOutput}
\end{CodeChunk}

\begin{figure}[!h]
    \centering
    \includegraphics[width=.7\textwidth]{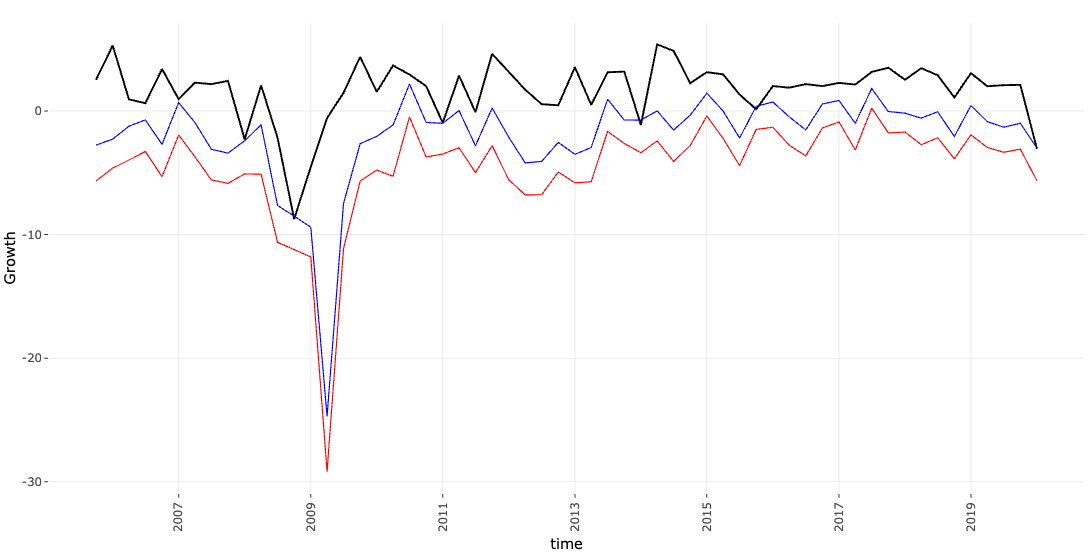}
    \caption{US quarterly growth: observed annualized rates (black), 1\% GaR (blue), and 1\% GiS stressed with $\alpha = 99\%$ (red).}
    \label{fig:gdp_final_plot}
\end{figure}

In Figure~\ref{fig:gdp_final_plot}, we plot the in-sample GaR and GiS estimates along with the dependent variables. As in \cite{Gonzlez-Rivera2024}, we observe that GiS is more negative than GaR. This negative outcome would be neglected if we only estimated GaR, which assumes that factors evolve according to an average scenario.

\section{Summary and discussion} \label{sec:summary}

The \pkg{FARS} package offers a suite of tools in \proglang{R} for modeling and designing economic scenarios based on conditional densities derived from MLDFMs and FA-QRs. It provides an integrated framework to extract multilevel factors with a flexible hierarchical structure, estimate their relationship with macroeconomic variables, and build stressed scenarios. The empirical applications presented in this paper illustrate how \pkg{FARS} can be used to analyze the distribution of European inflation and US GDP growth under adverse conditions, linking factor dynamics to macroeconomic outcomes. The \pkg{FARS} package is available on the Comprehensive R Archive Network (CRAN).






\section*{Acknowledgments}

\begin{leftbar}
Financial support from the Spanish Government grant PID2022-139614NB-C22/AIE/10.13039/501100011033 (MINECO/FEDER) is gratefully acknowledged by all authors. We are also very grateful to the editors and referees for their very constructive and useful comments, which helped to improve both the paper and the associated \pkg{FARS} package. Thanks are also due to Enrico Carraro, who, by using the \pkg{FARS} package, has been very helpful in finding some bugs, and to participants at the Computational and Financial Econometrics Conference (London, December 2025) and the Econometrics Seminar of the Statistics Department of the Universidad Carlos III de Madrid, for their comments. Part of this work was completed while
the third author was on sabbatical leave at Duke University. The author thanks the Department of Economics
for their hospitality. The third author thanks the support of the Asociación Mexicana de Cultura, A.C  Any remaining errors are the sole responsibility of the authors.
\end{leftbar}


\bibliography{refs}


\newpage

\begin{appendix}
\end{appendix}


\end{document}